%% file: tosem-main.tex
\documentclass[manuscript,nonacm]{acmart}

\AtBeginDocument{%
\providecommand\BibTeX{{%
\normalfont B\kern-0.5em{\scshape i\kern-0.25em b}\kern-0.8em\TeX}}}

\usepackage{booktabs} 
\usepackage{multirow}  
\usepackage{graphicx}
\usepackage{subcaption}
\usepackage{xcolor}

\usepackage{algorithm}
\usepackage{algpseudocode}
\usepackage{algorithmicx}
\usepackage{amsmath}

\usepackage{amssymb}
\usepackage{colortbl}
\usepackage{array}
\usepackage{mdframed}
\usepackage{tikz}
\usepackage{listings}
\usepackage{tikz}
\usetikzlibrary{positioning}

\usepackage{longtable}
\usepackage{booktabs}
\usepackage{enumitem}
\usepackage{makecell}
\usepackage[most]{tcolorbox} 
\tcbuselibrary{listings,skins,breakable}

\lstdefinestyle{customdiff}{
    basicstyle=\ttfamily\tiny\fontsize{6.5}{7.5}\selectfont,
    rulecolor=\color{black},
    tabsize=2,
    showspaces=false,
    showstringspaces=false,
    showtabs=false,
    breaklines=true,
    numbers=left,
    numberstyle=\tiny\fontsize{4}{5}\selectfont\color{gray},
    numbersep=2pt,
    xleftmargin=5pt,
    xrightmargin=1pt,
    framesep=1pt,
    framexleftmargin=7pt,
    captionpos=b,
    resetmargins=true,
    aboveskip=3pt,
    belowskip=3pt,
    gobble=0,
    escapeinside={(*@}{@*)},
    keywordstyle=\color{blue},
    commentstyle=\color{green!50!black},
    stringstyle=\color{orange},
    identifierstyle=\color{black},
    literate=
        {^+}{{{\color{DarkGreen}\bfseries+}}}1
        {^-}{{{\color{Sepia}\bfseries-}}}1,
    moredelim=[l][\color{DarkGreen}]{+},
    moredelim=[l][\color{Sepia}]{-},
    moredelim=[is][\color{blue}]{@@}{@@},
}

\definecolor{darkgreen}{rgb}{0.0, 0.5, 0.0}
\definecolor{darkred}{rgb}{0.6, 0.0, 0.0}
\definecolor{darkgray}{rgb}{0.5, 0.5, 0.5}
\definecolor{BrickRed}{RGB}{153, 51, 51}
\definecolor{Sepia}{RGB}{112, 66, 20}
\definecolor{DarkGreen}{RGB}{34, 100, 34}

\lstdefinelanguage{diff}{
  morecomment=[f][\color{red}]{-},
  morecomment=[f][\color{darkgreen}]{+},
  morecomment=[f][\color{darkgray}]{@},
}

\newboolean{showcomments}
\setboolean{showcomments}{true} 

\ifthenelse{\boolean{showcomments}}
 { \newcommand{\mynote}[2]{
      \fbox{\bfseries\sffamily\scriptsize#1}      {\small$\blacktriangleright$\textsf{\emph{#2}}$\blacktriangleleft$}}}
        { \newcommand{\mynote}[2]{}}

\def\pos#1{\textcolor{green!40!black}{+\,#1}}
\def\neg#1{\textcolor{red!50!black}{-\,#1}}

\newcommand{\toolname}[0]{\textit{Historian}}

\begin{document}

\title{Historian: Reducing Manual Validation in APR Benchmarking via Evidence-Based Assessment}



\author{Sahand Moslemi}
\email{sahand.moslemi@bilkent.edu.tr}
\affiliation{
  \institution{Bilkent University}
  \city{Ankara}
  \country{Turkey}
}

\author{Mayasah Lami}
\email{m.lami@bilkent.edu.tr}
\affiliation{
  \institution{Bilkent University}
  \city{Ankara}
  \country{Turkey}
}

\author{Anil Koyuncu}
\email{anil.koyuncu@cs.bilkent.edu.tr}
\orcid{0000-0001-6975-6752}
\affiliation{%
  \institution{Bilkent University}
  \city{Ankara}
  \country{Turkey}
}



\input{sections/abstract}

\ccsdesc[500]{Software and its engineering~Software maintenance tools}
\ccsdesc[500]{Software and its engineering~Software post-development issues}
\ccsdesc[500]{Software and its engineering~Software verification and validation}

\keywords{Automated Program Repair, Patch Correctness Assessment, Large Language Models, Code Clones, Software Engineering}



\input{sections/abstract}
\date{}
\maketitle

\input{sections/intro}

\input{sections/illustrate}

\input{sections/methodology}
\input{sections/setup}

\input{sections/evaluation-p1}

\input{sections/evaluation-p2}

\input{sections/evaluation-p3}
\input{sections/evaluation-p4}

\input{sections/evaluation-p5}

\input{sections/discussion}
\input{sections/threats}
\input{sections/related}

\input{sections/conclusion}


\bibliographystyle{ACM-Reference-Format}
\bibliography{ref}

\end{document}

%% file: sections/abstract.tex
\begin{abstract}

Assessing the correctness of patches generated by Automated Program Repair (APR) is a major bottleneck. Manual validation is labor-intensive and limited: exact matching overlooks valid variants, while semantic inspection is subjective and hard to reproduce. Existing Automated Patch Correctness Assessment (APCA) often relies on opaque predictive models that treat each patch as novel, repeatedly re-assessing semantically redundant patches. 

Our analysis of a large corpus of tool-generated patches reveals a duality: \textasciitilde39\% of unique correct patches are syntactic clones, suggesting opportunities for automation, yet \textasciitilde65\% of bugs have multiple distinct correct fixes, making single-reference assessment insufficient.

We present Historian, a framework that leverages Large Language Models to perform multi-reference comparisons against a knowledge base of historically validated patches, producing traceable, evidence-based verdicts while conservatively isolating novel cases as Unknown.

In leave-one-tool-out evaluation, Historian achieves 95.0\% coverage with 88.4\% accuracy, reducing manual validation to 5\% of patches. As an evidence-based pre-filter, enhancing the accuracy of standalone APCA tools by up to 21.8\% and enabling a hybrid pipeline with 86.2\% overall accuracy and 100\% coverage. A longitudinal analysis of tool-generated patches (2020–2024) shows that redundancy in repair attempts is common, indicating that many patches repeatedly rediscover established ones and strengthening the sustainability of evidence-based APR assessment.

\end{abstract}

%% file: sections/intro.tex
\section{Introduction}

APR has made significant strides in autonomously generating bug fixes. However, the ability to generate patches has far outpaced our ability to validate them. The central challenge in this domain is patch overfitting~\cite{qi2015analysis, long2016analysis, le2019reliability}, a phenomenon where a generated patch passes all tests, but fails to truly fix the underlying bugs or introduces unintended behaviors, compromising reliability.

Current evaluation practices typically rely on three methods: test suite augmentation~\cite{noda2020experience,humaninloop}, exact-match comparison against a developer's fix~\cite{hoppity,ueda2020devreplay}, and manual analysis~\cite{tbar,fixminer,jaid,coconut,Restore}. Each approach has limitations: test augmentation may fail to expose all behavioral faults due to incomplete coverage, exact matching ignores valid syntactic variants, and manual validation is costly, unscalable, and prone to bias~\cite{martinez2017automatic,yin2011fixes,martinez2018ultra}. Crucially, a fundamental limitation shared by these techniques is that they evaluate each patch in isolation, failing to leverage the vast knowledge embedded in the thousands of repairs already validated by the research community.

APCA techniques have emerged to address this bottleneck, spanning dynamic methods that execute patched programs~\cite{11-xin2017identifying,1-ye2021automated} and static approaches that analyze code features~\cite{6red-wang2020automated}. Despite their methodological diversity, many existing APCA tools are formulated as abstract predictive models that learn global features to estimate patch correctness. While such approaches are well suited and necessary for assessing repairs attempts for entirely novel bugs in production-centric environments, we argue that this predictive paradigm fails to capitalize on the historical cumulative knowledge inherent in benchmark-centric repair assessment in research settings. In this setting, researchers repeatedly assess the correctness of new tool-generated repair attempts for a fixed set of benchmark bugs (e.g., Defects4J). Treating every generated repair attempt as a novel prediction task is computationally inefficient and ignores a critical opportunity: many tool-generated repair attempts are semantically equivalent to with previously generated and thousands of repair attempts already validated by the research community. This systemic inefficiency leads to a bottleneck in benchmark-centric repair assessment, where human and computational resources are repeatedly wasted re-assessing recurring solutions.

Our work is grounded in a systematic empirical analysis that quantifies the scale of this unaddressed problem. By analyzing a large-scale corpus of tool-generated patches, we uncover a crucial duality: (1) a high degree of redundancy (\textasciitilde39\% of unique correct patches are syntactic clones), presenting a clear opportunity for automation; and (2) significant solution diversity (\textasciitilde65\% of bugs possess multiple, distinct correct fixes), indicating that any assessment methodology relying on a single reference patch is conceptually incomplete. This redundancy presents a critical opportunity: by automatically recognizing the semantic equivalences among repair attempts, we can eliminate a vast amount of repetitive validation effort and fundamentally change the economics of patch assessment in APR research.

This duality motivates Historian, a framework that shifts patch assessment from inferring correctness through opaque, model-driven predictions to evidence-based assessment grounded in traceable comparisons with historically validated patches. Rather than treating every candidate patch as a novel correctness prediction task, Historian builds on the observation that many tool-generated repairs are semantically equivalent to patches that have already been validated in prior peer-reviewed studies.



Historian operationalizes this vision by formulating correctness assessment as a multi-reference semantic equivalence detection problem. Leveraging the reasoning capabilities of large language models (LLMs), it performs exhaustive pairwise comparisons between a candidate patch and the Historical Reference Set to identify semantic equivalence. To translate these comparisons into reliable verdicts, Historian applies a two-stage evidence-based inference logic. First, it conducts pairwise inference to assign preliminary labels, Correct, Overfitting, or Unknown, based on the semantic relationship between the candidate and each reference. Second, it aggregates these labels using a consensus-based majority voting mechanism to produce a final correctness verdict. This design ensures that every high-confidence verdict is explicitly linked to a concrete peer-reviewed reference patch, making the assessment process transparent and verifiable. When no sufficiently strong evidence is found, Historian conservatively abstains by marking the repair attempt as Unknown. By intentionally prioritizing precision over coverage, the framework isolates genuinely novel repairs for expert inspection or subsequent analysis by complementary assessment techniques, while maintaining the evidential grounding of its decisions.



We evaluate Historian through a series of rigorous experiments validating its theoretical foundations and demonstrating its practical utility. In a 22-fold leave-one-tool-out evaluation, Historian autonomously determines the correctness 95.0\% of the patch dataset with 88.4\% accuracy. This high level of automation provides a significant reduction in manual validation workload by correctly identifying a vast majority of recurring repair attempts, thereby allowing researchers to focus their validation resources exclusively on the genuinely novel cases. Furthermore, we show that Historian acts as a robust pre-filter in hybrid pipelines, where Historian first renders correctness verdicts for repair attempts possessing historical precedent and a secondary APCA tool handles the residual Unknown cases, improving the accuracy of existing state-of-the-art models by up to 21.8\%. Finally, a longitudinal analysis (2020–2024) reveals that the recurrence of previously validated repair attempts is a persistent property of the benchmarking lifecycle, establishing a 21.8\% recurrence, including 40.0\% redundancy among correct patches, of previously validated patches. This finding confirms the long-term sustainability of the evidence-based paradigm, proving that the opportunity for automated assessment remains substantial and becomes increasingly powerful as the cumulative historical record grows.


We make the following contributions:
\begin{itemize}[leftmargin=*]

    \item We conduct and present the first large-scale empirical study that quantifies the duality of redundancy and diversity in APR generated patches, providing a foundation for our work.
    \item We propose Historian, a novel framework that introduces a new paradigm of traceable, evidence-based comparison for patch assessment.
    \item We conduct a large-scale evaluation of Historian's performance across diverse LLMs and configurations, measuring its coverage and accuracy as an automation tool.
    \item Through large-scale cross-tool validation, we demonstrate Historian's ability to reduce manual validation labor and its effectiveness in complementing and augmenting the accuracy of existing predictive APCA models.
    \item We perform a Sustainability Analysis via a five-year longitudinal study (2020–2024), demonstrating that the APR solution space is increasingly convergent; this provides an empirical mandate for Historian, as the potential for evidence-based assessment grows alongside the historical record.
    \item To promote transparency and facilitate the reproducibility of our findings, we have made all relevant artifacts publicly available at the following repository: 
    
    \url{https://anonymous.4open.science/r/Historian-Artifact}

\end{itemize}

\noindent \textbf{Our Vision:} Historian is not designed to replace predictive APCA tools or human experts, but to augment them. By filtering out the repetitive workload in APR benchmarking, Historian allows researchers to focus their attention on the novel repair attempts that truly advance the state of the art. By providing transparent, traceable verdicts linked to concrete historical evidence, Historian automates the validation of redundant solutions while enhancing the reliability of the entire patch assessment pipeline.

The remainder of this paper is organized as follows: Motivation (\S\ref{sec:motivation}), Methodology (\S\ref{sec:methodology}), Experimental Setup (\S\ref{sec:setup}), Evaluation (\S5), Discussion (\S\ref{sec:discussion}), Threats to Validity (\S\ref{sec:threats}), Related Work (\S\ref{sec:related_work}), and Conclusion (\S\ref{sec:conclusion}).

%% file: sections/illustrate.tex
\section{Motivation}
\label{sec:motivation}

Our work is motivated by a fundamental observation in APR: many tool-generated repair attempts appear textually unique but are semantically equivalent to previously validated patches. This redundancy suggests an untapped opportunity for automation; if a new repair attempt is semantically equivalent to a known validated patch, it may inherit that validation outcome, bypassing the need for repetitive manual inspection.

\noindent \textbf{Illustrative Example:} To illustrate the patch landscape, consider the Chart-24 bug from the Defects4J benchmark. Figure~\ref{fig:chart24-patches} presents three correct patches independently generated by various APR tools alongside the original human-written developer fix for this bug. These patches exhibit varying degrees of syntactic and semantic equivalence.

\input{tables/examples}

The patches generated by FixMiner~\cite{fixminer} and ssFix~\cite{xin2017leveraging} are textually identical, forming a Type-1 clone relationship. When compared to the developer's fix, these patches differ only in variable assignment style. This structural identity, combined with the use of different identifiers, represents a Type-2 clone relationship. In contrast, the Jaid~\cite{jaid} patch introduces a logically distinct conditional check. While it achieves the correct semantic outcome, its structural differences make it a Type-3 clone (partial similarity) relative to other patches.

These relationships suggest two critical insights into the patch assessment process. First, the prevalence of such code clone relationships implies that many "new" repair attempts may not be genuinely novel; rather, they could be syntactic or semantic rediscoveries of existing validated patches. If this pattern is pervasive, it suggests that a significant portion of manual validation effort is spent re-evaluating redundant outcomes.

Second, these examples suggest that relying solely on a single developer patch as the ground truth is methodologically narrow. This approach often obscures semantic equivalence: the fact that multiple "different" implementations can correctly fix the same bug. Under a strict exact-match or single-reference evaluation, a repair attempt that appears "different" from the developer fix might be erroneously flagged as incorrect, even when it is semantically equivalent. These observations motivate a shift toward a multi-reference ground truth, where every historically validated patch, whether authored by a human developer or an tool, could serve as a first-class reference point for assessment. By comparing a new repair attempt against an entire set of validated patches, we can assess its validity via its semantic equivalence to any previously established precedent.

\subsection{Empirical Motivation: Redundancy Among Correct Patches}

\label{sec:eval_theory}

\noindent \textbf{Approach:} To provide a rigorous empirical foundation for our work, we conduct a study to quantify two fundamental properties of the APR patch landscape: solution redundancy and solution diversity. By analyzing 327 tool-generated patches, we investigate whether the recurrence observed in our illustrative example generalizes across the broader ecosystem. These patches were selected from a merged dataset of five publicly available, peer-reviewed collections (detailed in Section~\ref{sec:patch-dataset}) and were previously validated as correct by the original study authors. To focus specifically on the diversity of alternative repair logic, we excluded any tool-generated patches that were exact textual matches to the original developer fix.

Our goal is to identify how many Solution Clusters, defined as groups of patches representing the same underlying conceptual fix, are contained within this set. By determining the distribution of these clusters, we can simultaneously measure redundancy (the extent to which different tools rediscover the same conceptual fix) and diversity (the number of distinct clusters, diverse implementations can correctly fix the same bug, associated with a single bug). To identify these clusters, we employ a four-stage progressive clustering pipeline where each subsequent stage merges clusters from the previous stage based on increasingly permissive definitions of equivalence:

\begin{enumerate}[leftmargin=*,nosep]
\item \textbf{Exact Match:} We first group patches whose modified method bodies are textually identical. This serves as our baseline, identifying only the most trivial redundancies.
\item \textbf{Token-based Match:} Utilizing SourcererCC\cite{sajnani2016sourcerercc} with a similarity threshold of 1.0, we identify Type-1 clones. This stage merges existing clusters, identified by exact match, that differ only in whitespace, comments, or minor formatting, which textual matching fails to capture.
\item \textbf{AST-based Match:} We apply an AST-based clone detection technique inspired by Deckard\cite{jiang2007deckard} to identify Type-2 clones. This stage further merges clusters that exhibit structural identity despite the use of different identifiers or literal values.
\item \textbf{Manual Assessment:} Finally, two authors manually inspected the remaining clusters to resolve subtle syntactic variants that automated tools may fail to capture. This included merging patches with "tool-brittle" variations, such as the addition of superfluous parentheses or minor formatting differences that do not alter the semantic outcome.
\end{enumerate}

\begin{figure}[ht]
    \includegraphics[width=0.72\columnwidth]{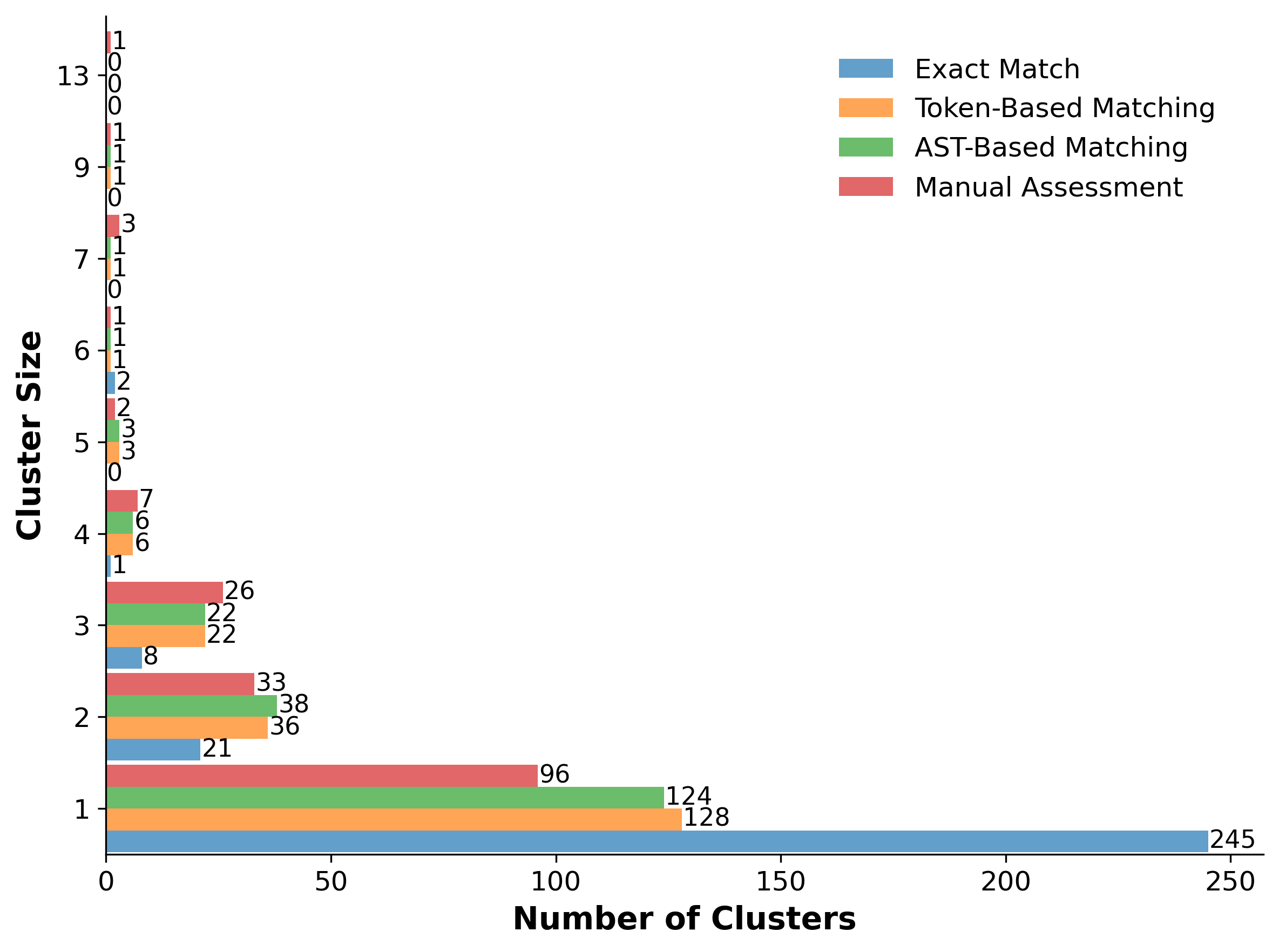}
    \caption{Distribution of Solution Clusters Under Different Equivalence Strategies.}
    \label{fig:dist-correct-bar}
\end{figure}

\noindent \textbf{Results:} Figure~\ref{fig:dist-correct-bar} illustrates the distribution of the 327 validated correct patches across clusters of varying sizes. To quantify redundancy, we categorize these clusters into two groups: (i) Unique Solutions (clusters of Size 1), representing a repair solution generated by only one tool in our dataset, and (ii) Recurring Solutions (clusters of Size 2+), representing a repair solution independently rediscovered by two or more tools.

The evidence for redundancy is found in the migration of patches from unique to recurring clusters as we incorporate syntactic and structural awareness. Under the Exact Match strategy, the dataset is highly fragmented: 245 of the 277 identified clusters consist of a single patch, suggesting that most tool-generated patches appear unique at a textual level. However, as we incorporate syntactic and structural awareness, these isolated patches merge. After Manual Assessment, the number of Size-1 clusters drops to 96, while the number of recurring clusters (Size 2 and above) increases significantly.

In total, 149 patches that appeared unique under exact textual matching were found to be clones of existing solutions. Overall, the total number of unique solutions is reduced from 277 to 170, representing a \textasciitilde39\% redundancy rate. This high degree of recurrence confirms that a substantial portion of manual validation in current APR research is spent re-evaluating Recurring Solutions, presenting a clear opportunity for automation.

While redundancy is high, the data also reveal substantial Solution Diversity, the fact that a single bug can be correctly fixed in multiple distinct ways. After manual assessment, the 170 unique solution clusters were mapped back to the 89 unique bugs from which they originated. Among these 89 bugs, 58 (\textasciitilde65\%) are associated with two or more distinct solution clusters even after manual inspection. This indicates that multiple correct fixes often coexist for the same bug, proving that a single-reference evaluation (matching only against the human developer fix) is conceptually insufficient to capture the full scope of valid repairs.

The 96 patches that remain in Size-1 clusters after manual assessment represent genuinely unique fixes within our current dataset. These patches are essential for building a comprehensive historical record. As repair research continues to evolve, these validated patches establish the initial reference points against which future repair attempts will be compared. Consequently, once a novel patch is validated and documented, it provides the necessary precedent to enable the automated assessment of any semantically equivalent attempts generated by future tools. The presence of these unique patches suggests that a historical record functions as a cumulative knowledge base, potentially expanding the scope of automated assessment as new and diverse repair outcomes are documented.

\begin{tcolorbox}[colback=blue!5,colframe=blue!50!black,title=Key Findings]
(1) There is significant solution redundancy: we observe a \textasciitilde39\% reduction in unique solutions after clustering patches that are syntactic clones (Type-1 and Type-2). This highlights a clear opportunity for automated workload reduction. \
(2) \textasciitilde65\% of bugs possess multiple semantically distinct correct solutions, proving that single-reference evaluation is methodologically narrow and motivating a multi-reference assessment paradigm.
\end{tcolorbox}

\subsection*{Implications}

Our empirical findings reveal a fundamental duality in the patch landscape: high solution redundancy among recurring repair attempts and significant solution diversity across different bug fixes. Together, these insights suggest two core design principles for an effective automated assessment framework.

\noindent \textbf{Multi-Reference Ground Truth.} The first principle is the necessity of a Multi-Reference Ground Truth. Because our study shows that approximately 65\% of bugs possess multiple semantically distinct correct fixes, an assessment framework that relies solely on a single human-written developer patch will inevitably suffer from a methodological blind spot. To accurately capture the full scope of semantic equivalence, a framework must treat every historically validated patch, whether authored by a human developer or a tool, as a first-class reference point. This approach ensures that a new repair attempt can be validated via its semantic equivalence to any known validated precedent, significantly expanding the available ground truth.

\noindent \textbf{Evidence-Based Inference Logic.} The second principle is the requirement for a principled Evidence-Based Inference Logic. While the observed redundancy confirms that identifying semantic equivalents can significantly reduce manual workload, such relationship alone is insufficient for a reliable assessment. A robust framework must provide a formal mechanism to translate the relationship between a new repair attempt and a historical reference into a definitive correctness verdict. This requirement stems from the inherent complexity of the patch space: for instance, a repair attempt may exhibit only partial similarity (e.g., a Type-3 clone) to a correct fix, which is insufficient to guarantee semantic equivalence. Conversely, an attempt could be a perfect clone of a known overfitting patch, necessitating a negative verdict despite its similarity to the historical record. Furthermore, the logic must distinguish between actionable evidence and genuine novelty: when a repair attempt lacks a clear semantic link to any validated precedent, the system should conservatively abstain from rendering a verdict. Therefore, a robust system cannot rely on naive similarity heuristics; it requires a set of principled rules to translate the available evidence, the specific clone type and the established label of the historical reference into a traceable correctness verdict. 

These principles form the conceptual foundation of the Historian, the framework proposed in this work. In the following sections, we describe how Historian operationalizes these principles by leveraging the semantic reasoning capabilities of LLMs to conduct multi-reference comparisons and provide traceable, evidence-based patch assessments.

%% file: tables/examples.tex
\lstdefinestyle{diffstyle}{
  style=customdiff,
  basicstyle=\ttfamily\scriptsize,
  escapeinside={(*@}{@*)},
  columns=fullflexible,
}

\newtcolorbox{codebox}[2][]{%
  colback=gray!10,
  colframe=gray!50,
  fonttitle=\bfseries,
  title=#2,
  listing only,
  listing options={style=diffstyle},
  #1
}

\begin{figure}[htbp]
    \centering
    
    \begin{tcolorbox}[
        enhanced,
        colback=white,
        colframe=blue!50!black,
        title=FixMiner \& ssFix Patches,
        fonttitle=\bfseries\footnotesize,
        left=0.5mm,
        right=0.5mm,
        top=0.5mm,
        bottom=0.5mm,
        width=0.95\columnwidth,
        boxsep=0pt,
        toptitle=0.5mm,
        bottomtitle=0.5mm,
        before=\vspace{-0.4mm},  
        after=\vspace{-0.4mm}
    ]
    \begin{lstlisting}[style=customdiff, basicstyle=\scriptsize\ttfamily]
@@ -122,7 +122,7 @@ public class GrayPaintScale
 public Paint getPaint(double value) {
  double v = Math.max(value, this.lowerBound);
- v = Math.min(v, this.upperBound);
+ value = Math.min(v, this.upperBound);
  int g = (int) ((value - this.lowerBound) / (this.upperBound - this.lowerBound) * 255.0);
  return new Color(g, g, g);
    \end{lstlisting}
    \end{tcolorbox}
    
    
    \begin{tcolorbox}[
        enhanced,
        colback=white,
        colframe=orange!50!black,
        title=Jaid Patch,
        fonttitle=\bfseries\footnotesize,
        left=0.5mm,
        right=0.5mm,
        top=0.5mm,
        bottom=0.5mm,
        width=0.95\columnwidth,
        boxsep=0pt,
        toptitle=0.5mm,
        bottomtitle=0.5mm,
        before=\vspace{-0.4mm},  
        after=\vspace{-0.4mm}
    ]
    \begin{lstlisting}[style=customdiff, basicstyle=\scriptsize\ttfamily]
@@ -123,6 +123,9 @@
 public Paint getPaint(double value) {
  double v = Math.max(value, this.lowerBound);
  v = Math.min(v, this.upperBound);
+ if((v == value) == false){
+  value=v;
+ }
  int g = (int) ((value - this.lowerBound) / (this.upperBound - this.lowerBound) * 255.0);
  return new Color(g, g, g);
    \end{lstlisting}
    \end{tcolorbox}
    
    
    \begin{tcolorbox}[
        enhanced,
        colback=white,
        colframe=green!50!black,
        title=Developer Patch,
        fonttitle=\bfseries\footnotesize,
        left=0.5mm,
        right=0.5mm,
        top=0.5mm,
        bottom=0.5mm,
        width=0.95\columnwidth,
        boxsep=0pt,
        toptitle=0.5mm,
        bottomtitle=0.5mm,
        before=\vspace{-0.4mm},  
        after=\vspace{-0.4mm}
    ]
    \begin{lstlisting}[style=customdiff, basicstyle=\scriptsize\ttfamily]
@@ -123,7 +123,7 @@ public class GrayPaintScale
 public Paint getPaint(double value) {
  double v = Math.max(value, this.lowerBound);
  v = Math.min(v, this.upperBound);
- int g = (int) ((value - this.lowerBound) / (this.upperBound 
+ int g = (int) ((v - this.lowerBound) / (this.upperBound 
        - this.lowerBound) * 255.0);
  return new Color(g, g, g);
 }
    \end{lstlisting}
    \end{tcolorbox}
    
    \caption{Comparison of repair patches for D4J Chart-24}
    \label{fig:chart24-patches}
\end{figure}

%% file: sections/methodology.tex
\section{Methodology}
\label{sec:methodology}

\begin{figure*}[ht]
    \centering
    \includegraphics[width=\textwidth]{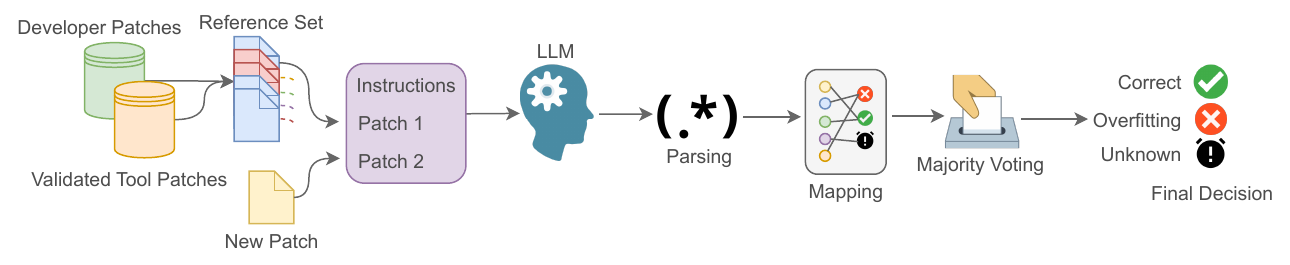}
    \vspace{-1.5em}
    \caption{Overview of \toolname}
    \label{fig:methodology-overview}
\end{figure*}

Figure~\ref{fig:methodology-overview} illustrates the architecture of Historian, which we design to automate the assessment of recurring repair attempts generated during APR benchmarking cycles. We reformulate patch evaluation as a multi-reference comparison task against a Historical Reference Set, which we construct by aggregating human-written developer fixes with previously validated tool-generated patches established in the literature. For each new candidate repair attempt, Historian leverages the reasoning capabilities of LLMs to conduct exhaustive pairwise comparisons against this historical record to determine whether the candidate is semantically equivalent to any previously validated patch. We process the resulting natural language responses of LLMs through a robust parsing pipeline to extract structured relationship labels that categorize the semantic link between the candidate and each reference (e.g., specific clone types). Finally, we translate these labels into correctness verdicts using our two-stage Evidence-Based Inference Logic. In the first stage, our Pairwise Inference assigns preliminary correctness verdicts to the candidate repair attempt by mapping its relationship to individual historical precedents. In the second stage, our Majority Voting aggregates these individual assessments into a single, high-confidence label. By anchoring every assessment in multi-reference comparison, our architecture facilitates the generation of correctness verdicts that are explicitly traceable back to the historical evidence contained within the reference set.

\subsection{Reference Set Construction}
\label{sec:reference_set}


To address the limitations of single-reference patch assessment and account for the solution diversity identified in Section~\ref{sec:motivation}, we construct a Historical Reference Set. This set incorporates the original developer patch alongside APR-generated patches that have been established as either Correct or Overfitting in prior peer-reviewed studies.

Formally, for a given bug $B$, we define the reference set $S_B$ as a collection of validated pairs:

\[
S_B = \{(P_{R_1}, L_1), (P_{R_2}, L_2), \dots, (P_{R_n}, L_n)\},
\] 

where $P_{R_i}$ represents a specific validated patch and $L_i \in \{\text{Correct, Overfitting}\}$ denotes its established ground-truth label. This multi-reference strategy serves as the foundation of our framework, enabling a more comprehensive evaluation by allowing new repair attempts to be assessed against a diverse body of historical evidence rather than a single point of reference. By including validated overfitting patches, we further enable the system to identify recurring repair errors, providing a more robust filter for automated assessment.

\subsection{Semantic Comparison and Prompting}
\label{sec:methodology-prompt-construction}



Given a Historical Reference Set $S_B$ for bug $B$, Historian evaluates each candidate repair attempt $P_C$ through exhaustive pairwise comparisons with every validated historical patch $P_{R_i} \in S_B$. For each pair $(P_C, P_{R_i})$, it leverages the semantic reasoning capabilities of LLMs to infer the nature of their relationship.


Deterministic code clone detection techniques, including textual (CCFinder~\cite{kamiya2002ccfinder}), token-based (SourcererCC~\cite{sajnani2016sourcerercc}), and AST-based (NiCad~\cite{cordy2011nicad}) approaches, are effective at identifying Type-1 and Type-2 clones. However, they struggle with Type-3 and Type-4 clones, where patches may diverge syntactically while preserving equivalent functional intent. Moreover, most traditional clone detectors operate as binary classifiers that only determine the presence of a clone relationship, without providing the fine-grained classification clone types (Type-1 through Type-4) required by our two-stage Evidence-Based Inference Logic.

To overcome these limitations, Historian leverages LLMs as semantic oracles. Recent empirical studies show that state-of-the-art LLMs achieve strong performance in semantic clone detection by reasoning about program intent and functional behavior rather than relying solely on surface-level syntactic similarity~\cite{dou2023towards, moumoula2025struggles}. By leveraging LLMs as semantic oracles, Historian generates structured relationship labels across multiple conceptual framings and contextual granularities. This multi-granular analysis provides the evidence necessary for the inference logic to differentiate levels of similarity, ranging from strict syntactic identity to deeper semantic equivalence, when rendering correctness verdicts.


To capture semantic relationships across multiple conceptual framings and contextual granularities, Historian performs three distinct classification tasks for each pair $(P_C, P_{R_i})$. We define these tasks to distinguish between syntactic structure and functional behavior, generating the relationship labels required for subsequent inference:

\begin{itemize}[leftmargin=*]
    \item \textbf{Code Clone Detection (CC):} A multi-class classification task that characterizes the relationship between two patches using the standard clone taxonomy. The model assigns a categorical label from the set  $ V_{\text{clone}} = \{\texttt{Type-1}, \texttt{Type-2}, \texttt{Type-3}, \texttt{Type-4}, \texttt{Not-a-Clone}\} $, ranging from strict syntactic duplication (Type-1) to semantic cloning (Type-4) achieved through divergent code structures.
    \item \textbf{Semantic Similarity (SS):} A binary classification task determining whether two patches implement overlapping functional behavior or similar repair intent, even if they are not fully behaviorally equivalent. The model provides a response from the set $V_{\text{binary}} = \{\texttt{yes}, \texttt{no}\}.$
    \item \textbf{Semantic Equivalence (SE):} A stricter binary classification task ($V_{\text{binary}}$) that assesses the semantic equivalence. It determines whether the candidate repair attempt and the validated patch are semantically equivalent with respect to the specific fault being repaired.
\end{itemize}

These notions form a hierarchy: semantic equivalence implies semantic similarity, while semantic similarity often, but not necessarily, corresponds to Type-3 or Type-4 clones. We systematically evaluate all three relationship framings across diverse prompting strategies and code representations to identify the configuration that provides the most reliable evidence.



To elicit the potential of LLM reasoning, Historian employs five distinct prompt designs adapted from prior work on LLM-based code clone detection~\cite{dou2023towards,moumoula2025struggles}. Table~\ref{tab:prompts} presents the complete prompts used in our study. We group these prompts into five high-level prompting strategies, summarized below.

\input{tables/prompts}


\begin{itemize}[leftmargin=*]
    \item \textbf{Simple (S):} Direct zero-shot prompts requesting a concise binary decision (yes/no) for semantic similarity or equivalence.
    \item \textbf{Reasoning-based (R):} One-step Chain-of-Thought (CoT) prompts that guide the model through a step-by-step reasoning process before producing a final yes/no label.
    \item \textbf{Line Similarity (LS):} Structure-aware prompts that first ask the model to compare lines between the two patches and report which lines correspond before determining semantic similarity or equivalence.
    \item \textbf{Simple Code Clone (SCC):} Prompts that require the model to explicitly classify the patch pair according to the standard clone types (Type-1, Type-2, Type-3, Type-4).
    \item \textbf{Integrated (I):} Multi-step prompts that guide the model through via detailed reasoning similarity scoring, clone type classification.
\end{itemize}




These prompting strategies provide complementary analytical lenses: while Simple prompts test immediate recognition, Reasoning-based prompts encourage deliberative analysis; Line Similarity prompts ground judgments in structural evidence, Simple Code Clone prompts invoke theoretical knowledge, and Integrated prompts synthesize multiple dimensions of comparison. 

Each prompt is instantiated using two complementary code representations to provide varying levels of contextual granularity: (i) the \textit{diff format}, which isolates added and removed lines and change introduced by the patch, and (ii) the \textit{full method body}, which provides broader contextual information including surrounding control flow and contextual information. While diffs aim to the focus attention on the change itself, full-method representations aim to enable reasoning about the patch’s impact on overall program semantics.

The resulting labels, spanning clone taxonomy, semantic similarity, and semantic equivalence, form the evidentiary basis for the inference and majority voting stages described in Section~\ref{sec:correctness-verdict}.

\subsection{LLM Response Parsing}
\label{sec:methodology-response-classification}


A persistent challenge with resource-constrained, small-scale open-source LLMs is their inconsistent adherence to rigid output format constraints~\cite{zeng2023evaluating, zhou2023instruction,xia2024fofo,chenBenchmarkingLargeLanguage2024}. Raw responses often incorporate conversational context, hedges, or auxiliary reasoning that deviates from the expected outputs~\cite{koyuncu2025exploring,macedoExploringImpactOutput2024}. While techniques such as supervised fine-tuning or controlled decoding~\cite{mudgal2023controlled} can improve structural consistency, they currently lack the capability to guarantee that generated outputs will invariably conform to user-defined constraints. Furthermore, recent research suggests that imposing rigid structural restrictions during the generation process can cause a significant decline in LLMs’ reasoning abilities, with stricter format constraints generally leading to greater performance degradation in complex tasks~\cite{tam2024let}.

To preserve the models’ reasoning integrity and ensure reliable downstream processing across diverse open-source models, we adopt a non-intrusive, two-stage extraction pipeline. Instead of forcing a strict output format, which could inadvertently suppress assessment accuracy, we adopt an architecture that makes no assumptions regarding the model's output format. We utilize a combination of regular expressions and zero-shot classification to implement a specialized parsing function $L(r, V)$ designed to extract a single well-defined relationship label from the raw response $r$. The function operates in two sequential stages:

\begin{enumerate}[leftmargin=*,nosep]
    \item \textbf{Regex Extraction:} We first attempt to identify a unique, case-insensitive keyword in $r$ that matches an element of the target set $V$.
    \item \textbf{Zero-Shot Fallback:} 
    If the regex extraction is ambiguous (i.e., it finds zero or multiple keywords), we employ a pre-trained zero-shot classification model. This model analyzes the full semantic context of the response $r$ to select the label from $V$ with the highest probability.
\end{enumerate}

Formally, we define the parsing function $L(r, V)$ as follows:

{\small
\begin{gather*}
L(r, V) =
\begin{cases}
\text{the unique element in } S, & \text{if } |S| = 1 \\[4pt]
\operatorname*{argmax}_{v \in V} P(v \mid r), & \text{otherwise}
\end{cases}
\end{gather*}
}

where $P(v \mid r)$ represents the semantic probability of label $v$ given the response $r$, as computed by the zero-shot classifier. The resulting relationship label is then utilized as the structured input for our subsequent correctness inference logic.

\subsection{Evidence-Based Inference Logic}

\label{sec:correctness-verdict}

The relationship labels generated during the parsing stage describe the semantic connection between two patches but do not constitute a direct correctness verdict. To translate this evidence into a final label for the candidate repair attempt, we utilize a two-stage process termed Evidence-Based Inference Logic.

\paragraph{1. Pairwise Inference}\label{sec:methodology-mapping}

In the first stage, we analyze each individual pair $(P_C, P_{R_i})$  to infer a preliminary correctness verdict for the candidate repair attempt $P_C$. We utilize the extracted relationship label $L(r, V)$ and the established ground-truth label $L_{i}$ of the reference patch to assign a preliminary status according to the rules detailed in Table~\ref{tab:translation}. We ground these rules in the following three technical rationales:

\input{tables/rules}
\begin{itemize}[leftmargin=*]
    \item \textbf{Semantic Equivalence:} If we identify a high-confidence relationship, specifically binary ``Yes'' labels (SS/SE) or taxonomic ``Type-1'', ``Type-2'', or ``Type-4'' clones, we assume $P_C$ is semantically equivalent to $P_{R_i}$.  In these instances, $P_C$ inherits the historical label $L_{i}$ (\emph{Correct} or \emph{Overfitting}). Type-2 clones are included because, in the APR context, candidate patches have already passed the test suite, so minor syntactic differences (e.g., identifiers or literals) are unlikely to alter the core functional logic.

    \item \textbf{Semantic Divergence:} We utilize explicit non-equivalence labels (`No` for SS/SE and `Not Clone` CC) to identify deviations from known solutions. If $P_C$ diverges from a Correct reference, we infer a preliminary verdict of Overfitting, as the repair attempt fails to implement a known-good solution. Conversely, if $P_C$ diverges from an Overfitting reference, we assign a label of Unknown, as divergence from a flawed repair provides no actionable evidence regarding the candidate's validity.

    \item \textbf{Structural Ambiguity:} We categorize Type-3 clones as Unknown. Because these represent only partial structural overlap, they provide insufficient evidence to guarantee semantic equivalence or correctness
\end{itemize}





\paragraph{2. Majority Voting for Final Verdict}\label{sec:methodology-voting}

In the second stage, we aggregate the multiple preliminary verdicts produced in the first stage to produce a single correctness verdict for the candidate $P_C$. We design this aggregation process to synthesize the evidence gathered across the entire Historical Reference Set, aiming to minimize the impact of potential reasoning errors within individual LLM-based comparisons.

To implement the voting mechanism, we first categorize the preliminary verdicts into informative votes (Correct or Overfitting) and uninformative votes (Unknown). We treat an Unknown verdict as an abstention, signaling that the LLM found insufficient evidence to establish a definitive relationship between the candidate and a specific historical reference. Because these uninformative votes represent a lack of actionable data, we exclude them from the final decision process. We adopt this approach based on the intuition that the identification of a clear semantic link to even a single validated patch constitutes sufficient grounds for a verdict, regardless of how many other comparisons remain inconclusive.



We then assign the final correctness verdict for $P_C$ as the majority class among the remaining informative votes. In the event of a tie between Correct and Overfitting labels, or if all pairwise comparisons result in Unknown (yielding no informative votes), we conservatively assign a final label of Unknown. This design choice is a core feature of our evidence-based paradigm, allowing us to intentionally trade coverage for higher precision and trustworthiness. In this context, trustworthiness is established by ensuring that the framework never forces a classification in the absence of verifiable precedent. By defaulting to Unknown when evidence is ambiguous or absent, we ensure that genuinely novel repair attempts, those lacking a validated historical precedent in our reference set, are isolated for expert review rather than being subjected to forced, potentially erroneous classification. Through this two-stage logic, we facilitate the generation of correctness verdicts that are explicitly traceable to verifiable historical evidence.

%% file: tables/prompts.tex
\begin{table*}[t]
\centering
\caption{Prompt Design for \toolname}
\label{tab:prompts}
\footnotesize
\renewcommand{\arraystretch}{1}

\setlength{\tabcolsep}{2pt}
\begin{tabular}{
>{\centering\arraybackslash}p{1.8cm}
>{\centering\arraybackslash}p{1.8cm}
>{\centering\arraybackslash}p{1.2cm}
|p{8cm}
}
\textbf{Task} & \textbf{Label} & \textbf{Strategy} & \textbf{Prompt} \\
\hline
\multirow{2}{*}{\textbf{CC}}  
 & \multirow{2}{*}{\textbf{Clone Type}} 
 & \textbf{SCC} & Please analyze the following two \textcolor{blue}{patches} / \textcolor{red}{code snippets} and determine if they are code clones. Respond with \textbf{‘yes’} if they are clones or \textbf{‘no’} if not. If \textbf{‘yes’}, please report the clone type: \textbf{Type-1}, \textbf{Type-2}, \textbf{Type-3}, or \textbf{Type-4}.\\
\cline{3-4}
 & & \textbf{I} & Please analyze the following two \textcolor{blue}{patches} / \textcolor{red}{code snippets} to assess their similarity and determine if they are code clones. Provide a similarity score between 0 and 10, where a higher score indicates more similarity. Additionally, identify the type of code clone they represent and present a detailed reasoning process for detecting code clones. \\
\hline
\multirow{3}{*}{\textbf{SS}} 
 & \multirow{3}{*}{\textbf{Yes/No}} 
 & \textbf{S} & Please analyze the following two \textcolor{blue}{patches} / \textcolor{red}{code snippets} and determine if they are semantically similar. Respond with \textbf{‘yes’} if they are similar or \textbf{‘no’} if not. \\
\cline{3-4}
 & & \textbf{R} & Please provide a detailed reasoning process for detecting semantic similarity in the following two \textcolor{blue}{patches} / \textcolor{red}{code snippets}. Based on your analysis, respond with \textbf{‘yes’} or \textbf{‘no’}. \\
\cline{3-4}
 & & \textbf{LS} & Please analyze the following two \textcolor{blue}{patches} / \textcolor{red}{code snippets} for semantic similarity. You should first report which lines of patches are more similar. Then, based on the report, please answer whether these two patches are semantically similar. The response should be \textbf{‘yes’} or \textbf{‘no’}. \\
\hline 
\multirow{3}{*}{\textbf{SE}} 
 & \multirow{3}{*}{\textbf{Yes/No}}  
 & \textbf{S} & Please analyze the following two \textcolor{blue}{patches} / \textcolor{red}{code snippets} and determine if they are semantically equivalent. Respond with \textbf{‘yes’} or \textbf{‘no’}. \\
\cline{3-4}
 & & \textbf{R} & Please provide a detailed reasoning process for detecting semantic equivalence in the following two \textcolor{blue}{patches} / \textcolor{red}{code snippets}. Based on your analysis, respond with \textbf{‘yes’} or \textbf{‘no’}. \\
\cline{3-4}
 & & \textbf{LS} & Please analyze the following two \textcolor{blue}{patches} / \textcolor{red}{code snippets} for semantic equivalence. You should first report which lines of patches are more similar. Then, based on the report, please answer whether these two patches are semantically equivalent. The response should be \textbf{‘yes’} or \textbf{‘no’}. \\
\end{tabular}
\end{table*}

%% file: tables/rules.tex
\begin{table}[t]
\centering
\caption{Pairwise Inference: Mapping Relationship Labels to Preliminary Verdicts}
\label{tab:translation}
\footnotesize

\setlength{\tabcolsep}{5pt}
\begin{tabular}{ccc|c}
\textbf{Task Type} & \textbf{Reference Patch Label} & \textbf{Relationship Label} & \textbf{Preliminary Verdict} \\
\hline
CC & Correct      & Not a Clone & \cellcolor{red!20}Overfitting \\
    \cline{3-4}

   &              & Type-1      & \cellcolor{green!20}Correct \\
    \cline{3-4}
   &              & Type-2      & \cellcolor{green!20}Correct \\
    \cline{3-4}
   &              & Type-3      & \cellcolor{yellow!20}Unknown \\
    \cline{3-4}
   &              & Type-4      & \cellcolor{green!20}Correct \\
\cline{2-4}
   & Overfitting  & Not a Clone & \cellcolor{yellow!20}Unknown \\
    \cline{3-4}
   &              & Type-1      & \cellcolor{red!20}Overfitting \\
    \cline{3-4}
   &              & Type-2      & \cellcolor{red!20}Overfitting \\
    \cline{3-4}
   &              & Type-3      & \cellcolor{yellow!20}Unknown \\
    \cline{3-4}
   &              & Type-4      & \cellcolor{red!20}Overfitting \\
\hline
SS / SE & Correct     & Yes & \cellcolor{green!20}Correct \\
    \cline{3-4}
        &             & No  & \cellcolor{red!20}Overfitting \\
\cline{2-4}
        & Overfitting & Yes & \cellcolor{red!20}Overfitting \\
    \cline{3-4}
        &             & No  & \cellcolor{yellow!20}Unknown \\
\end{tabular}
\end{table}

%% file: sections/setup.tex
\section{Experiment Setup}
\label{sec:setup}

\subsection{Research Questions}

We evaluate Historian through five research questions that investigate its theoretical foundations, implementation sensitivity, practical utility, and long-term sustainability:

\textbf{RQ1: What is the performance ceiling of the Historian framework?}
We conduct an oracle experiment to demonstrate the performance ceiling of Historian by simulating an ideal relationship inference. In this experiment, we replace the LLM with expert-annotated labels for the pairwise semantic relationships between each candidate repair attempt and the Historical Reference Set. This methodology allows us to evaluate our two-stage Evidence-Based Inference Logic (Pairwise Inference and Majority Voting) in isolation from the errors inherent in current LLMs. By establishing this analytical upper bound for both Coverage and Accuracy on the Covered Set, we validate whether our Evidence-Based Inference Logic accurately maps identified semantic relationships to definitive correctness verdicts.

\textbf{RQ2: How sensitive is Historian’s performance to its design choices?}
This research question investigates the impact of core design parameters, specifically LLM selection, prompting strategy, and code representation, on the trade-off between Coverage and Accuracy in the Covered Set. We first quantify the reliability of the two-stage response parser and its zero-shot fallback mechanism. We then perform a systematic sensitivity analysis to identify the configurations that optimize the balance between the volume of automated verdicts and the precision of the resulting labels.

\textbf{RQ3: To what extent does Historian reduce the manual validation workload in a large-scale benchmarking scenario?}
We evaluate Historian’s practical utility by simulating a large-scale APR benchmarking scenario through a 22-fold leave-one-tool-out cross-validation protocol. This experiment simulates the assessment of a new APR tool by treating the repair attempts of a single tool, representing a newly developed repair methodology, as the evaluation set, while dynamically reconstructing the Historical Reference Set from the validated patches of the remaining 21 tools. This setup represents the practical challenge where a researcher develops a new tool and must assess the correctness of its generated patches by leveraging the collective evidence established by existing repair methodologies. Our goal is to evaluate how well Historian functions as an evidence-based filter that can automatically determine the correctness of recurring repair attempts in a tool’s output. We assess this by measuring end-to-end Coverage and Accuracy on the Covered Set, demonstrating the framework’s ability to reduce the manual validation bottleneck across diverse repair tools.


\textbf{RQ4: To what extent can Historian serve as a filter to enhance the performance and efficiency of existing APCA tools?}
Historian acts as an evidence-based filter for redundant solutions, allowing assessment pipelines to bypass probabilistic prediction for previously validated repair attempts. This research question investigates the complementarity of a two-stage assessment pipeline, where Historian first assigns evidence-based correctness verdicts to recurring patches, enabling secondary predictive APCA tools to focus their inference exclusively on the genuinely novel residual cases. Utilizing a 22-fold leave-one-tool-out cross-validation protocol, we benchmark this integrated approach against three state-of-the-art predictive models. Our objective is to demonstrate that by delegating recurring repair attempts to an evidence-based framework, we can significantly enhance the accuracy and efficiency of existing assessment tools while providing a definitive correctness verdict for every patch in the dataset.

\textbf{RQ5: To what extent does solution redundancy persist as the APR benchmarking lifecycle evolves?}
We conduct a longitudinal redundancy analysis to investigate the empirical sustainability of evidence-based assessment. By chronologically evaluating tool-generated repair attempts produced between 2020–2024 for the Defects4J benchmark, we measure the frequency with which recurring solutions accumulate within the cumulative Historical Reference Set. To establish an objective lower bound for redundancy independent of LLM-specific bias, we utilize a deterministic, three-stage cascading detection pipeline consisting of: (i) textual identity check for modified method bodies, (ii) token-based similarity analysis via SourcererCC, and (iii) structural identity detection via AST-based tree-differencing. This analysis investigates whether the repetitive nature of the APR benchmarking lifecycle, in which new repair attempts for established bugs frequently yield semantically redundant solutions, provides a sustainable empirical mandate for a cumulative Historical Reference Set to mitigate the manual validation bottleneck.

\subsection{Patch Dataset}
\label{sec:patch-dataset}

Table~\ref{tab:merged-datasets} details the corpus of patches used across our experiments. We assemble a comprehensive evaluation corpus by merging five peer-reviewed, labeled Defects4J collections~\cite{patchsim, Liu_2020, 4-tian2020evaluating, 1-ye2021automated, 6red-wang2020automated}. The dataset comprises 1,455 unique tool-generated Correct patches and 37,858 unique Overfitting patches, covering 835 bugs from the Defects4J benchmark~\cite{Defects4J}.

To ensure the integrity of our evaluation corpus, we implement a three-stage sanitization process. First, we perform a structural validation of the unified diff files. We standardize malformed diff headers and discard any patch exhibiting persistent context misalignment, instances where the hunk metadata (line numbers and context snippets) fails to accurately map to the target file structure. Second, we conduct an application-level validation using git apply (v2.34.1, ignoring whitespace). We verify that each patch applies successfully to its corresponding Defects4J source code; any patch that fails this check is removed from the corpus. Finally, we de-duplicate the patches generated by the same tool for the same bug. We ensure that each patch generated by a specific tool for a given bug is unique, thereby preventing the artificial inflation of solution recurrence metrics and ensuring the statistical rigor of our subsequent analyses. Our resulting evaluation corpus focuses exclusively of single-method patches. We adopt this focus to ensure methodological tractability and to provide the LLM with a cohesive, self-contained control-flow context.

We prepare the sanitized tool-generated patches in two distinct formats to systematically evaluate the impact of contextual granularity on Historian’s assessment capacity (Section~\ref{sec:methodology-prompt-construction}). The diff format utilizes the original patch files to isolate the logical delta introduced by the repair. For the method format, we apply each patch to the source code and utilize the javalang library to parse the modified files, extracting the full source code of the altered function. This dual-representation strategy allows us to systematically evaluate the impact of contextual granularity on Historian’s ability to identify semantic equivalence.

\input{tables/datasets}

\subsection{Studied LLMS}

We evaluate Historian across a diverse suite of LLM), as summarized in Table~\ref{tab:llms}. We select these models based on following criteria: (\textit{i}) \textbf{Model Diversity}, including models from major code-focused families (CodeLlama, DeepSeek, Gemma, Qwen); (\textit{ii}) \textbf{Accessibility and Reproducibility}, prioritizing open-source models that are publicly available via Ollama (\texttt{v0.3.9}) framework to facilitate replication our results using accessible hardware ; and (\textit{iii}) \textbf{High-Resource Benchmarking}, for which we include the commercial Gemini 2.0 Flash model to provide a performance baseline for comparison against our open-source configurations.



\input{tables/llms}

We deployed all open-source Large Language Models using the Ollama framework (\texttt{v0.3.9}) on a local server equipped with 4 NVIDIA RTX 4090 GPUs (24GB VRAM each). This hardware configuration allowed us to run models with reasonable latency. For the commercial Gemini 2.0 Flash model, we utilized the Google AI Studio API.

\subsection{Zero-shot Classification}

As we established in our methodology (Section~\ref{sec:methodology-response-classification}), Historian requires a fallback mechanism to accurately interpret unstructured LLM responses. For this task, we utilize the \texttt{bart-large-mnli}~\footnote{https://huggingface.co/facebook/bart-large-mnli} model, a pre-trained Large Language Model optimized for Natural Language Inference (NLI). We utilize this model because its NLI approach, treating candidate labels as hypotheses to be evaluated against the raw LLM response as a premise, provides an established baseline for mapping the semantic intent of free-text responses to our structured label sets, $V_{binary}$ and $V_{clone}$~\cite{lewis2019bart}. By integrating this model into our parsing pipeline, we provide a mechanism to derive structured relationship labels from potentially complex or non-standard model outputs, thereby supporting the consistency of the subsequent inference stage.


\subsection{Evaluation Metrics}

Our evaluation reframes \toolname\ not as a traditional classifier, but as an automated evidence-based filter whose goal is to reduce manual validation workload. 
Accordingly, we adopt two primary metrics to quantify its performance: \textbf{Coverage} and \textbf{Accuracy on the Covered Set}.

Let $N$ be the total number of patches in an evaluation set. 
\toolname\ partitions this set into a Covered Set ($C$) comprising attempts assigned a definitive Correct or Overfitting label, and an Unknown Set ($U$) representing instances where the framework conservatively abstains from judgment. For the patches in the Covered Set ($C$), we use the standard definitions of True/False Positives (TP/FP) and True/False Negatives (TN/FN), where Correct is the positive class. We define our primary metrics as follows:

{\small
\begin{gather*}
\text{\textbf{Coverage}} = \frac{|C|}{N} = \frac{\text{TP} + \text{TN} + \text{FP} + \text{FN}}{N}, \text{\textbf{Accuracy on Covered Set}} = \frac{\text{TP} + \text{TN}}{|C|}
\end{gather*}
}


In this context, Coverage measures the proportion of the dataset that Historian can process autonomously, while Accuracy on the Covered Set  quantifies the reliability of those correctness verdicts. 

To evaluate the complementarity of Historian within the broader assessment ecosystem, we also utilize standard global Accuracy and F1-Score. These metrics are specifically applied to the hybrid assessment pipelines investigated in RQ4. Because these integrated architectures resolve the residual Unknown cases using secondary APCA tools, they provide a definitive correctness verdict for every repair attempt in the dataset. This allows us to quantify the performance enhancement achieved by utilizing Historian as an initial filtering stage that allows assessment pipelines to bypass probabilistic inference for previously validated repair attempts.


Finally, to validate the reliability of our manual expert annotations, we compute Cohen’s Kappa ($\kappa$) using the implementation in the \texttt{scikit-learn} library~\cite{scikit-learn}. This metric allows us to measure the degree of inter-rater agreement while inherently accounting for the probability of random agreement, ensuring the rigor of our oracle baseline.


%% file: tables/datasets.tex
\begin{table}[t]
  \caption{Dataset Statistics}
  \label{tab:merged-datasets}
  \centering
  \footnotesize   
  \begin{tabular}{p{3cm} c p{7cm}}
     & \textbf{B / C / O} & \textbf{Explanation} \\
    \midrule
    Xiong et al. \cite{patchsim} & 835 / 36 / 169 & \href{https://github.com/Ultimanecat/DefectRepairing}{\url{https://github/Ultimanecat/DefectRepairing}} \\ 
    Liu et al. \cite{Liu_2020} & 835 / 146 / 514 & \href{https://github.com/TruX-DTF/APR-Efficiency}{\url{https://github/TruX-DTF/APR-Efficiency}} \\ 
    Tian et al. \cite{4-tian2020evaluating} & 835 / 886 / 36,307 & \href{https://github.com/TruX-DTF/DL4PatchCorrectness}{\url{https://github/TruX-DTF/DL4PatchCorrectness}} \\ 
    Ye et al. \cite{1-ye2021automated} & 835 / 255 / 371 & \href{https://github.com/ASSERT-KTH/drr}{\url{https://github/ASSERT-KTH/drr}} \\ 
    Wang et al. \cite{6red-wang2020automated} & 835 / 132 / 497 & \href{https://zenodo.org/records/3730599}{\url{https://zenodo/records/3730599}} \\
    \midrule
    \textbf{Initial Collection} 
      & 835 / 1,455 / 37,858 
      & All Merged + Developer Patches \\

    \textbf{Applicable} 
      & 835 / 1,389 / 37,613 
      & Applicable Using \texttt{git apply} \\

    \textbf{Single-Method} 
      & 704 / 1,351 / 20,403 
      & Only Change One Method \\

    \textbf{De-duplicated} 
      & \textbf{704 / 450 / 8,704} 
      & Drop If Same Tool \& Same Target Method Content \\
    \midrule
    \multicolumn{3}{l}{\footnotesize \textbf{B:} Number of Bugs,\;
      \textbf{C:} Tool Correct Patches,\;
      \textbf{O:} Tool Overfitting Patches}
  \end{tabular}
\end{table}

%% file: tables/llms.tex
\begin{table}[t]
\centering
\caption{List of LLMs used in this study}
\label{tab:llms}
\footnotesize   

\setlength{\tabcolsep}{4pt}
\begin{tabular}{l l c c}
\textbf{Model UID} & \textbf{Description} & \textbf{Context Size} & \textbf{Abbr.} \\
\midrule
magicoder:7b-s-cl~\cite{wei2024magicoder}      & Magicoder 7B (Small)        & 8,192   & MC7B \\
codellama:7b-instruct~\cite{codellama2023code}& CodeLlama 7B               & 4,096   & CL7B \\
deepseek-coder:6.7b~\cite{deepseek2024}       & DeepSeek Coder 6.7B        & 16,000  & DSC6.7B \\
codegemma:7b-instruct~\cite{codegemma2024}    & CodeGemma 7B              & 32,768  & CG7B \\
qwen2.5:7b~\cite{bai2023qwen}                 & Qwen 2.5                  & 32,768  & QW7B \\
qwen2.5-coder:7b~\cite{qwencoder}             & Qwen 2.5 Coder            & 32,768  & QWC7B \\
yi-coder:9b~\cite{yi2024}                     & Yi Coder 9B               & 16,384  & YC9B \\
hermes3:8b~\cite{hermes2024}                  & Hermes 3                  & 8,192   & H3-8B \\
Gemini 2.0 Flash~\cite{gemini}                & Gemini 2.0 Flash          & Not disclosed & Gemini \\
\end{tabular}
\end{table}

%% file: sections/evaluation-p1.tex
\section{Evaluation}

\subsection{RQ1.  What is the performance ceiling of the Historian framework?}
\label{sec:RQ-oracle}

\textbf{Approach:} To validate the soundness of our framework’s core reasoning, we conduct an oracle experiment to establish its maximum performance potential. We simulate optimal conditions by replacing the LLM-based relationship detection with expert-annotated labels for the pairwise relationships between each candidate repair attempt and the Historical Reference Set. This allows us to evaluate our two-stage Evidence-Based Inference Logic (Pairwise Inference and Majority Voting) in isolation from the reasoning errors of current LLMs.

To simulate the assessment of unseen repairs, we use 139 repair attempts (40 Correct, 99 Overfitting) generated by the TBar repair tool as candidate repairs. These represent all TBar patches in our patch dataset (Section~\ref{sec:patch-dataset}) targeting the Defects4J benchmark. For each candidate, we construct a Historical Reference Set ($S_B$) that includes the original developer fixes and all other tool-generated patches previously validated as Correct or Overfitting. Crucially, to simulate a realistic cross-tool assessment scenario, we exclude all TBar patches from $S_B$, meaning the framework relies only on evidence from the developer fix or other tools’ patches to render a correctness verdict.


We utilize TBar for this manual analysis because its fix patterns were manually curated from the existing literature. Its generated patches therefore reflect repair strategies that have been repeatedly observed across prior tools. Furthermore, the scale of the TBar corpus permitted an exhaustive manual mapping of the entire pairwise relationship space, requiring 4,248 distinct expert annotations. While larger patch datasets exist, performing this level of dense, exhaustive annotation on a broader scale would be prohibitively labor-intensive. By labeling every possible relationship for these 139 candidates, we establish an oracle baseline that allows us to characterize the framework’s analytical upper bound.


To generate the oracle relationship labels, two authors, blinded to the ground-truth correctness of the candidates, independently categorized the 4,248 pairwise relationships based on the code clone taxonomy in Section~\ref{sec:methodology}. Disagreements occurred in only 136 instances (3.2\%), all of which we resolved through consensus discussion to reach a definitive set of oracle input. The high initial inter-annotator agreement (Cohen’s Kappa = 0.96) confirms the reliability and reproducibility of this human-derived input. For each of the 139 candidate repair attempts, we then executed our two-stage Evidence-Based Inference Logic: first applying the Pairwise Inference stage to translate the relationship labels into preliminary votes, followed by the Majority Voting stage to render final correctness verdicts.

\textbf{Results:} 

\textbf{From Pairwise Signals to Patch-Level Decisions.}
Figure~\ref{fig:rq1-oracle-pipeline} depicts the transformation of 4,248 manually annotated semantic relationships into definitive correctness verdicts through our three-step inference pipeline. Analysis of the expert-labeled pairs (Step~1) reveals substantial structural diversity within the APR solution space. Although textual identity (Type-1) and semantic equivalence across structurally divergent patches (Type-4) occur, over 86\% of relationships consist of partial similarity (Type-3) or complete divergence (Not-Clone). This uneven distribution highlights the difficulty of automated validation, as most individual comparisons do not provide decisive evidence and therefore require an inference logic capable of aggregating sparse signals.

\input{tables/p2-expert}


Despite this sparsity, Historian’s two-stage Evidence-Based Inference Logic is able to extract actionable evidence from the pairwise comparisons. In the Pairwise Inference stage (Step~2), the framework identifies 572 informative votes (275 Correct, 297 Overfitting), representing 13.5\% of all comparisons, while abstaining from 3,676 uninformative pairs.

The Majority Voting stage (Step~3) demonstrates the strength of the multi-reference strategy. Aggregating votes across the Historical Reference Set enables Historian to assign definitive correctness verdicts to 107 of 139 candidate repairs, achieving 77.0\% coverage. Thus, a sparse 13.5\% informative signal at the pairwise level translates to a 77.0\% patch-level assessment rate, indicating that multi-reference aggregation increases the likelihood of identifying semantically equivalent repairs.



The final confusion matrix (Step~3) characterizes the analytical soundness of the Historian framework. On the covered set, Historian achieved 100\% Accuracy, with zero instances of misclassifications. The remaining 32 patches (23.0\%) were correctly isolated as Unknown. Within our paradigm, these abstentions represent the successful isolation of truly novel repair attempts that lack verifiable historical precedent and thus require manual review.



\textbf{Illustrating the Evidence-Based Paradigm.}
To demonstrate the transparency of this process, we highlight the operational traces of two representative cases. Table~\ref{tab:cherry-Chart9TBarPatch497311} presents a successful validation for \texttt{Chart-9}. Despite the presence of 7 uninformative Unknown votes from divergent references, the logic identified 3 Type-2 clones matching previously validated patches from SketchFix and Elixir. This consensus of historical evidence enabled a Correct verdict. Table~\ref{tab:cherry-Closure109TBarpatch1} illustrates the detection of an Overfitting attempt for \texttt{Closure-109}. Here, the candidate was identified as a Type-1 clone of a known-bad attempt from \textit{kPAR}. Because the repair is a duplicate of an established error, the logic correctly inferred a status of Overfitting. These examples show that each Historian verdict is traceable and linked to verified patches.

\input{tables/p2-traces-correct}

\input{tables/p2-traces-overfitting}



\textbf{Stratified Analysis of Evidence Density.}
To characterize the robustness of the framework across varying levels of historical data, we performed a stratified analysis of Evidence Density, categorized into four strata. As summarized in Table~\ref{tab:evidence-density}, we categorized the 139 candidate repair attempts into four strata of decreasing evidence density, plus a final group for which no actionable evidence was identified. We define an Evidence Stratum as a grouping of repair attempts based on the number of informative votes supporting their final verdict.

\input{tables/stratified}
Analysis of the Single Precedent stratum (exactly one informative vote) reveals that Historian maintains perfect precision even when evidence is minimal. We observed that 27 repair attempts were assigned correctness verdicts based on a single semantic parallel, yet these verdicts matched the ground truth in 100\% of cases. This indicates that the framework's performance is driven by the specificity of the semantic link rather than the absolute volume of the Historical Reference Set. 

Conversely, for the 32 patches in the Strong and Moderate strata, Historian identified a robust consensus of evidence (averaging 14.8 and 6.2 informative votes, respectively). This high density provides a verifiability anchor, grounding the verdict in multiple independent repair attempts that have independently rediscovered the same logic. This redundancy facilitates a verifiable fact-checking process, allowing a researcher to audit an automated label by inspecting a diverse set of matched historical precedents.


The data further reveal that Historian successfully isolates these informative signals from high-volume, uninformative relationships. We define "Noise" as the uninformative pairwise verdicts (Type-3 or Not-Clone against overfitting references) that result in an Unknown preliminary vote. Across all resolved strata, the framework was exposed to significant noise, averaging 26.5 uninformative votes per patch. Even in the Single Precedent stratum, where the noise-to-signal ratio averaged 18:1, the framework correctly filtered out uninformative structural overlaps and rendered an accurate verdict.

\begin{tcolorbox}[colback=blue!5,colframe=blue!50!black,title=Summary of RQ1]
Achieving 100\% accuracy on the covered set proves that Historian’s inference logic is analytically sound, and clone-type relations are robust indicators of semantic equivalence. The results suggest that our inference logic has the potential to mitigate 77\% of the manual validation workload in the assessment of the TBar patches. These results establish a performance ceiling and demonstrate that evidence-based assessment provides a foundation for mitigating the manual validation workload in APR benchmarking.
\end{tcolorbox}

%% file: tables/p2-expert.tex
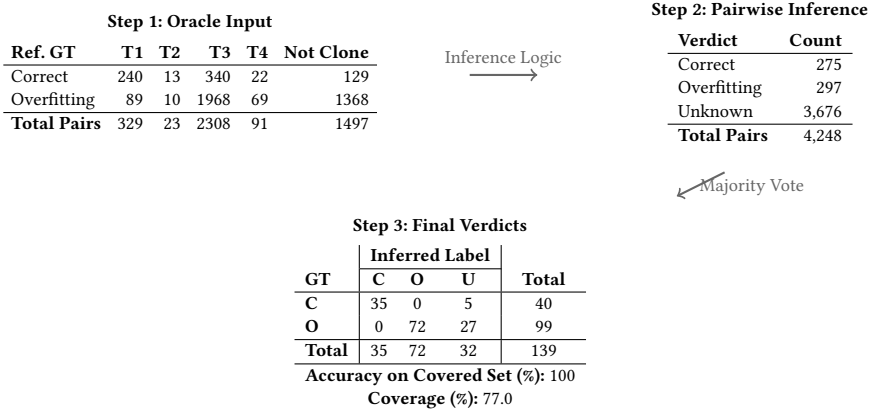
\begin{figure*}[t]
    \centering
    \caption{The RQ1 Oracle Experiment Pipeline: from Manual Annotations to Final Verdicts.}
    \label{fig:rq1-oracle-pipeline}
    \vspace{2mm}

    \footnotesize 

    \begin{tikzpicture}[node distance=0.9cm]

        \node (step1) {
            \begin{minipage}{0.48\linewidth}
                \centering
                \textbf{Step 1: Oracle Input}\\[1mm]
                \setlength{\tabcolsep}{3pt}
                \renewcommand{\arraystretch}{1.1}
                \begin{tabular}{lrrrrr}
                    \textbf{Ref. GT} & \textbf{T1} & \textbf{T2} & \textbf{T3} & \textbf{T4} & \textbf{Not Clone} \\
                    \hline
                    Correct      & 240 & 13 & 340  & 22 & 129 \\
                    Overfitting  & 89  & 10 & 1968 & 69 & 1368 \\
                    \hline
                    \textbf{Total Pairs} & 329 & 23 & 2308 & 91 & 1497 \\
                \end{tabular}
            \end{minipage}
        };

        \node[right=of step1] (step2) {
            \begin{minipage}{0.38\linewidth}
                \centering
                \textbf{Step 2: Pairwise Inference}\\[1mm]
                \setlength{\tabcolsep}{4pt}
                \renewcommand{\arraystretch}{1.1}
                \begin{tabular}{lr}
                    \textbf{Verdict} & \textbf{Count} \\
                    \hline
                    Correct      & 275 \\
                    Overfitting  & 297 \\
                    Unknown      & 3,676 \\
                    \hline
                    \textbf{Total Pairs} & 4,248 \\
                \end{tabular}
            \end{minipage}
        };

        \node[below=of step1, xshift=0.22\linewidth] (step3) {
            \begin{minipage}{0.62\linewidth}
                \centering
                \textbf{Step 3: Final Verdicts}\\[1mm]
                \setlength{\tabcolsep}{4pt}
                \renewcommand{\arraystretch}{1.1}
                \begin{tabular}{l|ccc|c}
                    & \multicolumn{3}{c|}{\textbf{Inferred Label}} & \\
                    \cline{2-4}
                    \textbf{GT} & \textbf{C} & \textbf{O} & \textbf{U} & \textbf{Total} \\
                    \hline
                    \textbf{C} & 35 & 0 & 5 & 40 \\
                    \textbf{O} & 0 & 72 & 27 & 99 \\
                    \hline
                    \textbf{Total} & 35 & 72 & 32 & 139 \\
                    \hline
                    \multicolumn{5}{c}{\textbf{Accuracy on Covered Set (\%):} 100} \\
                    \multicolumn{5}{c}{\textbf{Coverage (\%):} 77.0} \\
                \end{tabular}
            \end{minipage}
        };

        \draw[->, thick, gray!80!black]
            (step1.east) -- (step2.west)
            node[midway, above] {Inference Logic};


\draw[->, thick, gray!80!black, shorten >=100pt, shorten <=15pt]
    (step2.south) -- (step3.center)
    node[midway, right, xshift=35pt, yshift=18pt] {Majority Vote};

    \end{tikzpicture}

    \vspace{1mm}
    \parbox{0.98\linewidth}{\footnotesize \textbf{Note:} The figure shows the three stages of our RQ1 oracle experiment: the input distribution of manually annotated relationships (\textbf{Step 1}), the preliminary verdicts generated by our \textbf{Pairwise Inference} logic (\textbf{Step 2}), and the final patch-level confusion matrix after \textbf{Majority Voting} (\textbf{Step 3}). GT=Ground Truth, C=Correct, O=Overfitting, U=Unknown.}
\end{figure*}

%% file: tables/p2-traces-correct.tex
\begin{table}[t]
\centering
\caption{Cherry-Picked: Chart-9-TBar-Patch\_497\_311 (GT: Correct, Pred: Correct)}
\label{tab:cherry-Chart9TBarPatch497311}
\footnotesize
\renewcommand{\arraystretch}{1.1}
\setlength{\tabcolsep}{4pt}
\begin{tabular}{l|c|c|c}
\textbf{Reference Patch} & \textbf{Type} & \textbf{Ref. GT} & \textbf{Verdict} \\
\hline
Chart-9-SketchFix-patch1 & T2 & C & \textcolor{green!60!black}{C} \\
Chart-9-Elixir-patch1 & T2 & C & \textcolor{green!60!black}{C} \\
Chart-9-developer & T2 & C & \textcolor{green!60!black}{C} \\
Chart-9-Nopol-DifferentFragPatches-0 & NC & O & \textcolor{gray}{U} \\
Chart-9-Nopol2017-Patch88 & NC & O & \textcolor{gray}{U} \\
Chart-9-Jaid-patch3 & NC & O & \textcolor{gray}{U} \\
Chart-9-Nopol-patch1 & NC & O & \textcolor{gray}{U} \\
Chart-9-Jaid-patch2 & T3 & O & \textcolor{gray}{U} \\
Chart-9-Jaid-patch1 & T3 & O & \textcolor{gray}{U} \\
Chart-9-SequenceR-patch1 & T3 & C & \textcolor{gray}{U} \\
\hline
\multicolumn{3}{r|}{\textbf{Verdict Counts:}} & C:3, U:7 \\
\multicolumn{3}{r|}{\textbf{Final Prediction:}} & \textbf{Correct} \\
\hline
\end{tabular}
\vspace{1mm}
\parbox{\linewidth}{\centering\footnotesize 
\vspace{1mm}
\textbf{Type:} T1-T4: Type-1 to Type-4, NC: Not-Clone. \;
\textbf{Ref. GT:} C: Correct, O: Overfitting. \\
\textbf{Verdict:} \textcolor{green!60!black}{C}: Correct, \textcolor{gray}{U}: Unknown.}
\end{table}


%% file: tables/p2-traces-overfitting.tex
\begin{table}[t]
\centering
\caption{Cherry-Picked: Closure-109-TBar-patch1 (GT: Overfitting, Pred: Overfitting)}
\label{tab:cherry-Closure109TBarpatch1}
\footnotesize
\renewcommand{\arraystretch}{1.1}
\setlength{\tabcolsep}{4pt}
\begin{tabular}{l|c|c|c}
\textbf{Reference Patch} & \textbf{Type} & \textbf{Ref. GT} & \textbf{Verdict} \\
\hline
Closure-109-kPAR-Patch\_172\_38 & T1 & O & \textcolor{orange}{O} \\
Closure-109-Dynamoth-DifferentFilePatches-0 & NC & O & \textcolor{gray}{U} \\
Closure-109-SimFix-Patch\_6\_6 & NC & O & \textcolor{gray}{U} \\
Closure-109-developer & T3 & C & \textcolor{gray}{U} \\
\hline
\multicolumn{3}{r|}{\textbf{Verdict Counts:}} & O:1, U:3 \\
\multicolumn{3}{r|}{\textbf{Final Prediction:}} & \textbf{Overfitting} \\
\hline
\end{tabular}
\vspace{1mm}
\parbox{\linewidth}{\centering\footnotesize 
\vspace{1mm}
\textbf{Type:} T1-T4: Type-1 to Type-4, NC: Not-Clone. \;
\textbf{Ref. GT:} C: Correct, O: Overfitting. \\
\textbf{Verdict:} \textcolor{orange}{O}: Overfitting, \textcolor{gray}{U}: Unknown.}
\end{table}

%% file: tables/stratified.tex
\begin{table}[h]
\centering
\caption{Stratified Analysis of Evidence Density and Verdict Accuracy (n=139)}
\label{tab:evidence-density}
\footnotesize
\renewcommand{\arraystretch}{1.1}
\begin{tabular}{l|c|c|c|c|c}
\textbf{Evidence Strata} & \textbf{Count} & \textbf{Avg. Pairs} & \textbf{Avg. Informative} & \textbf{Avg. Noise} & \textbf{Accuracy} \\
\textit{(Informative Votes)} & \textbf{(Patches)} & \textbf{(Pairs)} & \textbf{(T1, T2, T4, NC)} & \textbf{(T3, NC)} & \textbf{(Covered)} \\
\hline
\textbf{Strong Consensus ($\ge$ 10)} & 11 & 56.4 & 14.8 & 41.6 & 100\% \\
\textbf{Moderate Evidence (5--9)} & 21 & 35.8 & 6.2 & 29.6 & 100\% \\
\textbf{Sparse Evidence (2--4)} & 48 & 31.4 & 2.8 & 28.6 & 100\% \\
\textbf{Single Precedent (1)} & 27 & 19.3 & 1.0 & 18.3 & 100\% \\
\hline
\textbf{No Actionable Evidence (0)} & 32 & 23.5 & 0.0 & 23.5 & N/A  \\
\hline
\textbf{Total / Average} & \textbf{139} & \textbf{30.6} & \textbf{4.1} & \textbf{26.5} & \textbf{100\%} \\
\end{tabular}
\end{table}

%% file: sections/evaluation-p2.tex
\subsection{RQ2. How sensitive is Historian’s performance to its design choices?} \label{sec: RQ2}
\label{sec:RQ-design}

\textbf{Approach:} This research question evaluates the practical implementation of Historian, focusing on how its performance is influenced by three primary design parameters: (i) the underlying LLM, (ii) the prompting strategy, and (iii) the code representation (Diff vs. Method). Unlike the oracle study, which established an analytical performance ceiling, this experiment reflects a realistic benchmarking context where semantic relationships are inferred by the models themselves. We utilize the 139 TBar patches as our evaluation set to perform a systematic sensitivity analysis across 128 unique configurations (8 models x 8 prompts x 2 representations).

The evaluation proceeds in two stages. First, we assess the reliability of the two-stage response parser by measuring the frequency of successful regex extraction and the accuracy of the zero-shot fallback. Second, we analyze the trade-off between Coverage and Accuracy on the Covered Set across all configurations to identify the optimal parameters for subsequent large-scale experiments.

\textbf{Results:}



\textbf{Instruction Adherence and Formatting Robustness.}
Accurate extraction of structured relationship labels is a prerequisite for downstream inference. Relying solely on prompt-based formatting instructions proves insufficient to guarantee output consistency across models. We observed a significant disparity in Instruction Adherence: as shown in Table~\ref{tab:regex_accuracy}, the average regex extraction rate across all tasks was 53.0\%, with significant variance. On average, QW7B followed the requested format in 74.0\% of cases, MC7B succeeded in only 19.0\% of instances. Furthermore, the data suggest that adherence is sensitive to task complexity and representations. For example, the QW7B model achieved high adherence on binary tasks (avg. 0.80 for SS/SE) but dropped to an average of 0.23 for multi-class clone classification (CC) when using the diff representation. This collective inconsistency across models, tasks, and representations reinforces our architectural decision to utilize a format-agnostic extraction pipeline that interprets model responses without assuming a structured output.

\input{tables/zeroshot-adherence}

\textbf{Automated Validation of the Zero-Shot Fallback.}
We first assess the Zero-Shot Fallback on responses where regex extraction succeeds, treating regex labels as ground truth. Across all models, the fallback recovers intended labels with 90.0\% average accuracy (Table~\ref{tab:zeroshot_accuracy}), demonstrating strong reconstruction under structured conditions.

\input{tables/zeroshot-accuracy}

\textbf{Manual Validation of the Zero-Shot Fallback.}
To ensure this reliability extends to the unstructured responses where the regex fails, we performed an extensive manual analysis by sampling 20 responses for each of the 128 experimental strata, resulting in 2,560 manually verified samples. As detailed in Table~\ref{tab:manual_zeroshot_accuracy}, the results reveal that model families such as QW7B, QWC7B, and H8B exhibit the highest coherence, with manual parsing accuracies ranging from 0.91 to 0.95. This suggests that these models produce highly consistent traces that are accurately classified by the zero-shot fallback. We observed that the Simple Code Clone (SCC) prompt strategy yielded the most reliable parsing results (avg. 0.96). In contrast, the Integrated (I) prompt proved the most challenging to parse (avg. 0.56), a result we attribute to the increased complexity of requiring the model to generate both a numerical similarity score and a categorical classification within a single response. We hypothesize that this multi-objective framing may obscure the model's primary intent, making it more difficult for the zero-shot classifier to identify the relationship label. Notably, several configurations achieved 100\% manual accuracy, indicating that the zero-shot fallback can reliably map unstructured natural language explanations to relationship labels.


\textbf{Sensitivity Analysis: Accuracy on the Covered Set.}
As summarized in Table~\ref{tab:accuracy}, Historian achieved an overall average accuracy of 0.84, demonstrating its capacity to provide assessments under varying configurations.


\input{tables/p2-comparison-accuracy}

The results reveal that performance is highly sensitive to the interaction between model and prompting strategy. The highest observed accuracy (0.94) was achieved by the QW7B model using the Simple Code Clone (SCC) prompt on Diff-level representations. However, no single prompt strategy was universally optimal. While SCC was the best performer for the Qwen and CodeLlama families, models like YC9B and CG7B reached their peak performance using Reasoning-based (R) or Integrated (I) prompts.

\textbf{Sensitivity Analysis: Automation Coverage.}
While accuracy measures the reliability of the rendered correctness verdicts, Coverage (Table~\ref{tab:coverage}) reflects the breadth of applicability. It quantifies the proportion of repair attempts for which Historian can establish a semantic link to validated historical patches. Historian achieved a high average coverage of 94.8\% across all tested configurations.

\input{tables/p2-comparison-coverage}

The primary factor influencing coverage is the code representation. Method-level representations consistently yielded higher coverage, achieving an average of 96.1\% compared to 92.6\% for the diff format. We attribute this to the full method body providing a more complete control-flow and data-flow context, allowing the model to identify semantic links more frequently. 



As summarized in Table~\ref{tab:accuracy} and Table~\ref{tab:coverage}, while several Method-level configurations achieved slightly higher coverage (e.g., reaching 97\% to 99\%), they consistently exhibited lower accuracy (ranging from 0.80 to 0.88). We select the QW7B model using the SCC prompt on the diff representation as the optimal configuration, as it achieved the highest observed accuracy of 0.94 while maintaining a coverage of 95.0\%.

Our results indicate a trade-off centered on code representation. Method-level context maximizes the volume of automated assessments. In contrast, the diff representation yields higher peak accuracy. We hypothesize that the more focused context of the diff representation constrains the model to base its judgment primarily on the core code changes. This reduces false associations from surrounding code, leading to more abstentions but higher precision.


Comparing these results with the RQ1 oracle study (100\% Accuracy, 77.0\% Coverage) reveals a 18\% coverage gap. Historian’s optimal configuration (QW7B, SCC prompt, Diff) achieves 0.94 Accuracy with 95.0\% Coverage, exceeding the oracle’s coverage while remaining below its 100\% accuracy. This indicates that LLMs are capable of identifying valid semantic patterns in complex cases, where human annotators, following the strict manual protocol utilized in RQ1 leads to more frequent abstentions.



\begin{tcolorbox}[colback=blue!5,colframe=blue!50!black,title=Summary of RQ2]
Historian’s optimal configuration (\textbf{QW7B, SCC prompt, Diff}) achieves 0.94 Accuracy with95.0\% Coverage. This performance exhibits high alignment with the precision of the human-annotated oracle (100\% Acc) while significantly expanding automation volume beyond the oracle’s 77.0\% Coverage. Beyond overall performance, we identify a trade-off: Method-level representations maximize the quantity of verdicts (Coverage), while Diff-level representations enable peak precision (Accuracy). Furthermore, we find that our two-stage response parser facilitates the generation of accurate relationship labels from unstructured LLM outputs.
\end{tcolorbox}

%% file: tables/zeroshot-adherence.tex
\begin{table*}[t]
\centering
\caption{Instruction Adherence}
\label{tab:regex_accuracy}
\footnotesize
\setlength{\tabcolsep}{2pt}
\renewcommand{\arraystretch}{1.1}
\begin{tabular}{l|ccc|ccc|ccc|ccc|cc|cc|c}
\multirow{4}{*}{\textbf{Model}}
& \multicolumn{6}{c|}{\textbf{SS (Yes/No)}}
& \multicolumn{6}{c|}{\textbf{SE (Yes/No)}}
& \multicolumn{4}{c|}{\textbf{CC (No/Clone-Type)}}
& \multirow{4}{*}{\textbf{Avg.}} \\
\cline{2-7}\cline{8-13}\cline{14-17}
& \multicolumn{3}{c|}{\textbf{Method}} & \multicolumn{3}{c|}{\textbf{Diff}}
& \multicolumn{3}{c|}{\textbf{Method}} & \multicolumn{3}{c|}{\textbf{Diff}}
& \multicolumn{2}{c|}{\textbf{Method}} & \multicolumn{2}{c|}{\textbf{Diff}} & \\
\cline{2-4}\cline{5-7}\cline{8-10}\cline{11-13}\cline{14-15}\cline{16-17}
& S & R & LS & S & R & LS & S & R & LS & S & R & LS & SCC & I & SCC & I & \\
\hline
MC7B & 0.07 & 0.28 & \cellcolor{blue!15}0.39 & 0.06 & 0.16 & 0.26 & 0.04 & 0.27 & 0.29 & 0.08 & 0.23 & 0.33 & 0.23 & 0.11 & 0.10 & 0.10 & 0.19 \\
\hline
CL7B & 0.49 & 0.21 & 0.24 & 0.73 & 0.21 & 0.18 & 0.52 & 0.20 & 0.24 & \cellcolor{blue!15}0.92 & 0.16 & 0.18 & 0.60 & 0.17 & 0.33 & 0.06 & 0.34 \\
\hline
DSC6.7B & 0.34 & 0.38 & 0.31 & 0.52 & 0.63 & 0.54 & 0.27 & 0.50 & 0.32 & 0.51 & \cellcolor{blue!15}0.70 & 0.64 & 0.39 & 0.22 & 0.54 & 0.13 & 0.43 \\
\hline
CG7B & 0.80 & 0.56 & 0.70 & 0.96 & 0.87 & 0.49 & 0.81 & 0.57 & 0.78 & 0.95 & 0.86 & 0.95 & 0.66 & 0.09 & \cellcolor{blue!15}0.97 & 0.01 & 0.69 \\
\hline
QW7B & 0.69 & 0.69 & 0.91 & 0.96 & 0.89 & 0.94 & 0.56 & 0.76 & 0.88 & \cellcolor{blue!15}0.97 & 0.93 & 0.96 & 0.61 & 0.61 & 0.24 & 0.23 & 0.74 \\
\hline
QWC7B & 0.97 & 0.66 & 0.80 & \cellcolor{blue!15}0.98 & 0.78 & 0.69 & 0.97 & 0.65 & 0.83 & \cellcolor{blue!15}0.98 & 0.82 & 0.65 & 0.63 & 0.60 & 0.21 & 0.15 & 0.71 \\
\hline
YC9B & 0.61 & 0.67 & 0.63 & 0.53 & 0.81 & 0.61 & 0.53 & 0.68 & 0.56 & 0.60 & \cellcolor{blue!15}0.85 & 0.59 & 0.46 & 0.10 & 0.14 & 0.11 & 0.53 \\
\hline
H8B & 0.68 & 0.60 & 0.56 & 0.93 & 0.74 & 0.33 & 0.39 & 0.55 & 0.50 & \cellcolor{blue!15}0.95 & 0.81 & 0.53 & 0.72 & 0.46 & 0.94 & 0.19 & 0.62 \\
\hline
\textbf{Avg.} & 0.58 & 0.51 & 0.57 & 0.71 & 0.63 & 0.50 & 0.51 & 0.52 & 0.55 & 0.74 & 0.67 & 0.60 & 0.54 & 0.30 & 0.43 & 0.12 & 0.53 \\
\bottomrule
\end{tabular}
\parbox{\linewidth}{\centering\footnotesize \colorbox{blue!15}{Blue} indicates the best value(s) for each model.}
\end{table*}



%% file: tables/zeroshot-accuracy.tex
\begin{table*}[t]
\centering
\caption{Zero-Shot Classification Accuracy}
\label{tab:zeroshot_accuracy}
\footnotesize
\setlength{\tabcolsep}{2pt}
\renewcommand{\arraystretch}{1.1}
\begin{tabular}{l|ccc|ccc|ccc|ccc|cc|cc|c}
\multirow{4}{*}{\textbf{Model}}
& \multicolumn{6}{c|}{\textbf{SS}}
& \multicolumn{6}{c|}{\textbf{SE}}
& \multicolumn{4}{c|}{\textbf{CC}}
& \multirow{4}{*}{\textbf{Avg.}} \\
\cline{2-7}\cline{8-13}\cline{14-17}
& \multicolumn{3}{c|}{\textbf{Method}} & \multicolumn{3}{c|}{\textbf{Diff}}
& \multicolumn{3}{c|}{\textbf{Method}} & \multicolumn{3}{c|}{\textbf{Diff}}
& \multicolumn{2}{c|}{\textbf{Method}} & \multicolumn{2}{c|}{\textbf{Diff}} & \\
\cline{2-4}\cline{5-7}\cline{8-10}\cline{11-13}\cline{14-15}\cline{16-17}
& S & R & LS & S & R & LS & S & R & LS & S & R & LS & SCC & I & SCC & I & \\
\hline
MC7B &0.96 &0.96 & \cellcolor{blue!15}0.98 &0.93 &0.94 & \cellcolor{blue!15}0.98 &0.97 & \cellcolor{blue!15}0.98 &0.96 &0.94 &0.94 &0.96 &0.82 &0.73 &0.66 &0.74 &0.90 \\
\hline
CL7B &0.95 &0.68 &0.87 & \cellcolor{blue!15}1.00 &0.90 &0.97 &0.95 &0.69 &0.89 & \cellcolor{blue!15}1.00 &0.85 &0.98 & \cellcolor{blue!15}1.00 &0.66 &0.99 &0.83 &0.89 \\
\hline
DSC6.7B &0.94 &0.96 &0.87 &0.97 & \cellcolor{blue!15}0.98 &0.97 &0.93 & \cellcolor{blue!15}0.98 &0.91 &0.97 & \cellcolor{blue!15}0.98 &0.97 &0.84 &0.64 &0.96 &0.71 &0.91 \\
\hline
CG7B &0.99 &0.98 &0.99 & \cellcolor{blue!15}1.00 & \cellcolor{blue!15}1.00 & \cellcolor{blue!15}1.00 &0.99 &0.98 &0.99 & \cellcolor{blue!15}1.00 & \cellcolor{blue!15}1.00 & \cellcolor{blue!15}1.00 & \cellcolor{blue!15}1.00 &0.52 & \cellcolor{blue!15}1.00 &0.62 &0.94 \\
\hline
QW7B &0.94 &0.87 &0.91 & \cellcolor{blue!15}1.00 &0.92 &0.91 &0.92 &0.86 &0.90 & \cellcolor{blue!15}0.99 &0.86 &0.90 & \cellcolor{blue!15}0.99 &0.54 &0.95 &0.67 &0.88 \\
\hline
QWC7B &0.95 &0.87 &0.93 & \cellcolor{blue!15}1.00 &0.96 & \cellcolor{blue!15}1.00 &0.94 &0.87 &0.93 & \cellcolor{blue!15}1.00 &0.97 & \cellcolor{blue!15}1.00 & \cellcolor{blue!15}1.00 &0.37 & \cellcolor{blue!15}1.00 &0.77 &0.91 \\
\hline
YC9B &0.97 &0.98 &0.99 & \cellcolor{blue!15}1.00 & \cellcolor{blue!15}1.00 &0.99 &0.97 &0.99 &0.97 &0.98 & \cellcolor{blue!15}1.00 & \cellcolor{blue!15}1.00 &0.77 &0.63 &0.20 &0.89 &0.90 \\
\hline
H8B &0.97 &0.91 &0.48 & \cellcolor{blue!15}1.00 &0.98 &0.96 &0.94 &0.83 &0.56 & \cellcolor{blue!15}1.00 &0.98 &0.98 & \cellcolor{blue!15}1.00 &0.44 & \cellcolor{blue!15}1.00 &0.72 &0.86 \\
\hline
\textbf{Avg.} &0.96 &0.90 &0.88 &0.99 &0.96 &0.97 &0.95 &0.90 &0.89 &0.98 &0.95 &0.97 &0.93 &0.57 &0.84 &0.74 &0.90 \\
\bottomrule
\end{tabular}
\vspace{1mm}
\parbox{\linewidth}{\centering\footnotesize \colorbox{blue!15}{Blue} indicates the best value(s) for each model.}
\end{table*}

\begin{table*}[t]
\centering
\caption{Zero-Shot Classification Accuracy on Manually Evaluated Samples}
\label{tab:manual_zeroshot_accuracy}
\footnotesize
\setlength{\tabcolsep}{2pt}
\renewcommand{\arraystretch}{1.1}
\begin{tabular}{l|ccc|ccc|ccc|ccc|cc|cc|c}
\multirow{4}{*}{\textbf{Model}}
& \multicolumn{6}{c|}{\textbf{SS (Yes/No)}}
& \multicolumn{6}{c|}{\textbf{SE (Yes/No)}}
& \multicolumn{4}{c|}{\textbf{CC (No/Clone-Type)}}
& \multirow{4}{*}{\textbf{Avg.}} \\
\cline{2-7}\cline{8-13}\cline{14-17}
& \multicolumn{3}{c|}{\textbf{Method}} & \multicolumn{3}{c|}{\textbf{Diff}}
& \multicolumn{3}{c|}{\textbf{Method}} & \multicolumn{3}{c|}{\textbf{Diff}}
& \multicolumn{2}{c|}{\textbf{Method}} & \multicolumn{2}{c|}{\textbf{Diff}} & \\
\cline{2-4}\cline{5-7}\cline{8-10}\cline{11-13}\cline{14-15}\cline{16-17}
& S & R & LS & S & R & LS & S & R & LS & S & R & LS & SCC & I & SCC & I & \\
\hline
MC7B & 0.80 & 0.65 & 0.75 & 0.50 & 0.50 & 0.80 & 0.95 & 0.55 & 0.60 & 0.50 & 0.60 & 0.70 & \cellcolor{blue!15}1.00 & 0.50 & 0.95 & 0.90 & 0.70 \\
\hline
CL7B & 0.90 & 0.75 & 0.55 & 0.45 & 0.55 & 0.75 & 0.80 & 0.75 & 0.65 & \cellcolor{blue!15}1.00 & 0.65 & 0.95 & 0.95 & 0.70 & 0.70 & 0.50 & 0.73 \\
\hline
DSC6.7B & 0.80 & 0.80 & 0.85 & 0.40 & 0.70 & 0.65 & 0.80 & \cellcolor{blue!15}1.00 & \cellcolor{blue!15}1.00 & 0.35 & 0.60 & 0.90 & 0.90 & 0.50 & 0.55 & 0.35 & 0.70 \\
\hline
CG7B & 0.95 & 0.80 & \cellcolor{blue!15}1.00 & 0.95 & 0.70 & 0.85 & 0.95 & 0.80 & \cellcolor{blue!15}1.00 & 0.90 & 0.95 & 0.90 & 0.95 & 0.60 & 0.95 & 0.70 & 0.87 \\
\hline
QW7B & \cellcolor{blue!15}1.00 & \cellcolor{blue!15}1.00 & \cellcolor{blue!15}1.00 & \cellcolor{blue!15}1.00 & 0.95 & 0.95 & \cellcolor{blue!15}1.00 & 0.95 & 0.95 & \cellcolor{blue!15}1.00 & 0.95 & \cellcolor{blue!15}1.00 & \cellcolor{blue!15}1.00 & 0.90 & \cellcolor{blue!15}1.00 & 0.50 & 0.95 \\
\hline
QWC7B & \cellcolor{blue!15}1.00 & \cellcolor{blue!15}1.00 & 0.95 & \cellcolor{blue!15}1.00 & 0.95 & 0.90 & \cellcolor{blue!15}1.00 & 0.95 & \cellcolor{blue!15}1.00 & \cellcolor{blue!15}1.00 & \cellcolor{blue!15}1.00 & 0.85 & \cellcolor{blue!15}1.00 & 0.80 & \cellcolor{blue!15}1.00 & 0.55 & 0.93 \\
\hline
YC9B & 0.75 & 0.90 & \cellcolor{blue!15}1.00 & 0.95 & 0.90 & 0.90 & 0.80 & 0.95 & 0.90 & 0.60 & 0.70 & \cellcolor{blue!15}1.00 & 0.85 & 0.85 & \cellcolor{blue!15}1.00 & 0.60 & 0.85 \\
\hline
H8B & \cellcolor{blue!15}1.00 & 0.95 & 0.95 & \cellcolor{blue!15}1.00 & \cellcolor{blue!15}1.00 & \cellcolor{blue!15}1.00 & \cellcolor{blue!15}1.00 & \cellcolor{blue!15}1.00 & 0.95 & \cellcolor{blue!15}1.00 & 0.85 & 0.90 & \cellcolor{blue!15}1.00 & 0.65 & \cellcolor{blue!15}1.00 & 0.35 & 0.91 \\
\hline
\textbf{Avg.} & 0.90 & 0.86 & 0.88 & 0.78 & 0.78 & 0.85 & 0.91 & 0.87 & 0.88 & 0.79 & 0.79 & 0.90 & 0.96 & 0.69 & 0.89 & 0.56 & 0.83 \\
\bottomrule
\end{tabular}
\vspace{1mm}
\parbox{\linewidth}{\centering\footnotesize \colorbox{blue!15}{Blue} indicates the best value(s) for each model.}
\end{table*}

%% file: tables/p2-comparison-accuracy.tex
\begin{table*}[t]
\centering
\caption{Accuracy Under Covered Set}
\label{tab:accuracy}
\footnotesize
\setlength{\tabcolsep}{2pt}
\renewcommand{\arraystretch}{1.1}
\begin{tabular}{l|ccc|ccc|ccc|ccc|cc|cc|c}
\multirow{4}{*}{\textbf{Model}}
& \multicolumn{6}{c|}{\textbf{SS}}
& \multicolumn{6}{c|}{\textbf{SE}}
& \multicolumn{4}{c|}{\textbf{CC}}
& \multirow{4}{*}{\textbf{Avg.}} \\
\cline{2-7}\cline{8-13}\cline{14-17}
& \multicolumn{3}{c|}{\textbf{Method}} & \multicolumn{3}{c|}{\textbf{Diff}}
& \multicolumn{3}{c|}{\textbf{Method}} & \multicolumn{3}{c|}{\textbf{Diff}}
& \multicolumn{2}{c|}{\textbf{Method}} & \multicolumn{2}{c|}{\textbf{Diff}} & \\
\cline{2-4}\cline{5-7}\cline{8-10}\cline{11-13}\cline{14-15}\cline{16-17}
& S & R & LS & S & R & LS & S & R & LS & S & R & LS & SCC & I & SCC & I & \\
\hline
MC7B   & 0.84 & 0.82 & 0.82 & 0.80 & 0.75 & 0.72 & 0.85 & 0.86 & 0.81 & 0.82 & 0.86 & 0.72 & \cellcolor{blue!15}\textbf{0.87} & 0.82 & 0.75 & 0.85 & 0.81 \\
\hline
CL7B   & 0.85 & 0.78 & 0.81 & 0.73 & 0.81 & 0.75 & 0.88 & 0.84 & 0.84 & 0.67 & 0.85 & 0.76 & \cellcolor{blue!15}\textbf{0.90} & 0.89 & 0.85 & 0.81 & 0.81 \\
\hline
DSC6.7B & 0.80 & 0.84 & 0.82 & 0.86 & 0.85 & 0.82 & 0.77 & 0.83 & 0.82 & 0.80 & 0.87 & 0.82 & 0.84 & 0.82 & 0.79 & \cellcolor{blue!15}\textbf{0.87} & 0.83 \\
\hline
CG7B   & 0.86 & 0.84 & 0.89 & 0.85 & 0.87 & 0.83 & 0.85 & 0.86 & 0.89 & 0.86 & \cellcolor{blue!15}\textbf{0.89} & 0.84 & 0.88 & 0.88 & 0.85 & 0.85 & 0.86 \\
\hline
QW7B   & 0.86 & 0.89 & 0.85 & 0.92 & 0.89 & 0.92 & 0.88 & 0.89 & 0.86 & 0.85 & 0.90 & 0.85 & 0.91 & 0.88 & \cellcolor{green!20}\textbf{0.94} & 0.85 & 0.88 \\
\hline
QWC7B  & 0.85 & 0.86 & 0.83 & 0.88 & 0.92 & 0.90 & 0.85 & 0.85 & 0.86 & 0.88 & 0.86 & 0.81 & 0.85 & 0.83 & \cellcolor{blue!15}\textbf{0.93} & 0.87 & 0.86 \\
\hline
YC9B   & 0.85 & 0.88 & 0.90 & 0.82 & 0.89 & 0.76 & 0.91 & 0.89 & 0.90 & 0.76 & 0.82 & 0.78 & 0.83 & 0.87 & 0.73 & \cellcolor{blue!15}\textbf{0.92} & 0.84 \\
\hline
H8B    & 0.87 & 0.85 & 0.83 & 0.87 & \cellcolor{blue!15}\textbf{0.90} & 0.86 & 0.86 & 0.88 & 0.86 & 0.73 & 0.86 & 0.77 & 0.81 & 0.84 & 0.90 & 0.89 & 0.85 \\
\hline
\textbf{Avg.} & 0.85 & 0.85 & 0.84 & 0.84 & 0.86 & 0.82 & 0.86 & 0.86 & 0.86 & 0.80 & 0.86 & 0.79 & 0.86 & 0.85 & 0.84 & 0.86 & 0.84 \\
\bottomrule
\end{tabular}
\vspace{1mm}
\parbox{\linewidth}{\centering\footnotesize \colorbox{blue!15}{Blue} indicates the best value for each model. \colorbox{green!20}{Green} indicates the best performance.}
\end{table*}

%% file: tables/p2-comparison-coverage.tex
\begin{table*}[t]
\centering
\caption{Coverage (\%) Under Various Configurations}
\label{tab:coverage}
\footnotesize
\setlength{\tabcolsep}{2pt}
\renewcommand{\arraystretch}{1.1}
\begin{tabular}{l|ccc|ccc|ccc|ccc|cc|cc|c}
\multirow{4}{*}{\textbf{Model}}
& \multicolumn{6}{c|}{\textbf{SS}}
& \multicolumn{6}{c|}{\textbf{SE}}
& \multicolumn{4}{c|}{\textbf{CC}}
& \multirow{4}{*}{\textbf{Avg.}} \\
\cline{2-7}\cline{8-13}\cline{14-17}
& \multicolumn{3}{c|}{\textbf{Method}} & \multicolumn{3}{c|}{\textbf{Diff}}
& \multicolumn{3}{c|}{\textbf{Method}} & \multicolumn{3}{c|}{\textbf{Diff}}
& \multicolumn{2}{c|}{\textbf{Method}} & \multicolumn{2}{c|}{\textbf{Diff}} & \\
\cline{2-4}\cline{5-7}\cline{8-10}\cline{11-13}\cline{14-15}\cline{16-17}
& S & R & LS & S & R & LS & S & R & LS & S & R & LS & SCC & I & SCC & I & \\
\hline
MC7B   & 97.1 & 97.8 & 96.4 & 94.2 & 90.6 & 93.5 & \cellcolor{blue!15}98.6 & 95.7 & 92.8 & 96.4 & 94.9 & 92.1 & 91.7 & 94.2 & 92.8 & 96.4 & 94.7 \\
\hline
CL7B   & 98.6 & 96.4 & 93.5 & 89.2 & 94.2 & 92.8 & \cellcolor{blue!15}99.3 & 97.1 & 96.4 & 86.3 & 94.9 & 92.8 & 96.4 & 97.8 & 94.9 & 97.1 & 94.9 \\
\hline
DSC6.7B & 91.4 & 89.9 & 88.5 & 91.4 & 92.1 & 97.1 & 92.1 & 91.4 & 92.1 & 91.4 & 95.7 & \cellcolor{blue!15}97.8 & 92.1 & 93.5 & 88.5 & 97.1 & 92.6 \\
\hline
CG7B   & 97.1 & 96.4 & 96.4 & \cellcolor{blue!15}98.6 & \cellcolor{blue!15}98.6 & 95.7 & 97.8 & 95.7 & 94.2 & 97.8 & 97.1 & 92.1 & 97.1 & 96.4 & 87.8 & 97.8 & 96.0 \\
\hline
QW7B   & \cellcolor{blue!15}97.8 & 95.7 & 96.4 & 95.0 & 96.4 & \cellcolor{blue!15}97.8 & \cellcolor{blue!15}97.8 & \cellcolor{blue!15}97.8 & 97.1 & 92.8 & 95.0 & 92.8 & 96.3 & 94.2 & 95.0 & \cellcolor{blue!15}97.8 & 96.0 \\
\hline
QWC7B  & \cellcolor{blue!15}97.8 & 95.7 & 95.7 & 92.1 & 97.1 & 94.2 & \cellcolor{blue!15}97.8 & \cellcolor{blue!15}97.8 & 95.7 & 90.6 & 95.7 & 91.4 & \cellcolor{blue!15}97.8 & 97.1 & 92.1 & 97.1 & 95.4 \\
\hline
YC9B   & \cellcolor{blue!15}98.6 & 97.1 & 97.8 & 87.9 & 95.3 & 89.7 & \cellcolor{blue!15}98.6 & 95.7 & 97.8 & 87.9 & 94.4 & 86.0 & 94.2 & \cellcolor{blue!15}98.6 & 83.2 & 92.5 & 93.5 \\
\hline
H8B    & 97.8 & 97.8 & 97.8 & 95.0 & 95.7 & 95.7 & 97.8 & \cellcolor{blue!15}98.6 & 95.7 & 86.3 & 95.7 & 95.7 & 96.4 & 97.1 & 90.6 & 95.7 & 95.6 \\
\hline
\textbf{Avg.} & 97.0 & 95.9 & 95.3 & 92.9 & 95.0 & 94.6 & 97.5 & 96.2 & 95.2 & 91.2 & 95.4 & 92.6 & 95.3 & 96.1 & 90.6 & 96.4 & 94.8 \\
\bottomrule
\end{tabular}
\vspace{1mm}
\parbox{\linewidth}{\centering\footnotesize \colorbox{blue!15}{Blue} indicates the best value for each model.}
\end{table*}

%% file: sections/evaluation-p3.tex
\subsection{RQ3. To what extent does Historian reduce the manual validation workload in a large-scale benchmarking scenario?}
\label{sec:RQ-Effectiveness}

\textbf{Approach:}
We evaluate the practical utility of Historian in a large-scale benchmarking scenario, simulating the assessment of repair attempts. Unlike conventional APCA tools that rely on abstract predictive models, Historian utilizes its Historical Reference Set as a knowledge base for evidence-based assessment. To rigorously test the Historian's capacity to generalize to novel tool outputs, we adopt the 22-fold leave-one-tool-out (LOTO) cross-validation protocol established in prior work~\cite{zhouLeveragingLargeLanguage2024}.


Our evaluation utilizes a corpus of 825 repair attempts generated by 22 distinct tools. To ensure that our findings remain comparable to recent APCA literature, we adopt the standardized patch dataset compiled by Lin et al.~\cite{Cache}. While some repair tools are reported to have generated larger sets of patches in their original publications (e.g., TBar), we utilized the specific subset as reported in the baseline study\cite{Cache}. 

In each experimental fold, repair attempts from one tool are held out as an unseen test set. To maintain a strict separation of evidence and prevent data leakage, we dynamically reconstruct the Historical Reference Set for each fold, populating it exclusively with the validated patches established by the remaining 21 tools. This design assesses two critical dimensions of the framework: (i) its capacity to generalize to the unique repair signatures of an unseen tool architecture, and (ii) its effectiveness in identifying recurring solutions across a heterogeneous landscape of repair methodologies. Furthermore, this protocol allows us to evaluate Historian’s performance across a broad spectrum of historical evidence. Because we construct the reference set  $S_B$ on a per-bug basis, the LOTO experiment inherently assesses the framework's reliability even for bugs with limited historical data. Based on our findings in RQ2, we utilize the QW7B configuration (SCC prompt, Diff representation) as our primary open-source baseline and compare it against the commercial Gemini 2.0 Flash model to analyze how the framework’s assessment capacity scales with increased model model size.

\textbf{Results:} 

\textbf{Overall Performance: Coverage and Accuracy.}  
The results of the leave-one-tool-out cross-validation, detailed in Table~\ref{tab:gemini-comparison}, demonstrate that Historian provides a substantial reduction in manual workload. We report the weighted average across all 22 folds, where each fold’s contribution is proportional to the number of repair attempts generated by the respective tool. Using the open-source QW7B model, Historian achieved a weighted average Coverage of 92.5\% with an Accuracy on the Covered Set of 84.7\%. The Gemini 2.0 Flash configuration improved the weighted average results to 95.0\% Coverage and 88.4\% Accuracy.

\input{tables/rq3-internal}


\textbf{Impact on Manual Validation Workload.}  
These results quantify the framework's practical impact: in a realistic scenario, the Gemini configuration successfully assigned definitive correctness verdicts to 784 out of 825 total repair attempts (95.0\% of the aggregate corpus). Additionally, in 88.4\% of the cases, the framework automates the vast majority of the validation process with high reliability. This high level of coverage directly translates to workload reduction: instead of manually validating 825 patches, a research team utilizing Historian can focus its limited resources exclusively on the 41 instances (5.0\%) that the framework conservatively marked as \emph{Unknown}. 


\textbf{Robustness Across APR Tool Families.}  
The results further indicate that Historian’s assessment capacity is not strictly limited by the generating tool’s underlying repair methodology, provided that semantically equivalent repairs are present in the cumulative historical record. The framework demonstrated robust performance across a diverse range of architectures, reaching near-perfect accuracy and total coverage for several tools. For instance, using the Gemini configuration, Historian achieved accuracies exceeding 93\% for tools as diverse as Arja, CapGen, Cardumen, GenProg, and Nopol. This consistency suggests that, despite substantial methodological differences among APR systems, many repair attempts converge toward semantically equivalent patches that Historian can recognize independent of their origin.

Furthermore, the high performance across all 22 folds establishes Historian’s robustness to varying reference set densities. Although the size of the reference set varies according to the historical availability of validated repairs in Defects4J, the framework maintained high precision even for bugs with limited historical data. This outcome indicates that the framework’s effectiveness depends less on the absolute volume of prior patches and more on its ability to identify semantically equivalent patches. This observation aligns with our findings in RQ1, where we showed that a small number of informative votes is often sufficient to determine a candidate’s correctness.

\textbf{Analysis of Outlier Tool Performance.}  
While Historian performs well across most tool families, the accuracy for ACS and SOFix decreases to 59\% and 50\%, respectively. Unlike the other 20 tools in our study, ACS and SOFix exhibit Correct-dominant repair attempts (e.g., a 29:8 ratio for ACS and 10:1 for SOFix). The lower accuracy observed for these tools indicates that the framework rendered incorrect verdicts rather than conservatively abstaining. This suggests that repair attempts from these tools possess characteristics that make them particularly susceptible to false semantic equivalence detection within our evidence-based logic. These findings are consistent with existing APCA benchmarks, where predictive models also demonstrate a marked performance decrease when evaluating ACS and SOFix. These outliers underscore that for certain repair methodologies, human-in-the-loop verification remains a necessary component to ensure the overall reliability of the assessment pipeline.


\textbf{Evidence-Based Verifiability and Benchmarking Implications.}  
The practical implication of these results is a shift in the validation workflow of APR benchmarking. Out of the 825 total repair attempts, only 41 (5.0\%) were flagged as Unknown, representing the small fraction of repair logic that is truly novel or semantically isolated from the historical record. For each of the 784 repair attempts, Historian establishes a traceable link between the candidate and the specific historical precedents. This ensures that the automated assessment is a transparent conclusion grounded in verified historical evidence. This coupling of volume reduction and evidence-linked reliability provides a scalable solution for making the APR benchmarking lifecycle more efficient.




\begin{tcolorbox}[colback=blue!5,colframe=blue!50!black,title=Summary of RQ3]
Historian facilitates a 95.0\% reduction in manual validation workload by autonomously assigning correctness verdicts to 784 out of 825 repair attempts with 88.4\% accuracy (using Gemini 2.0 Flash). This performance validates the framework’s ability to reliably automate the assessment of redundant repair attempts in APR benchmarking, significantly mitigating the manual validation bottleneck. Furthermore, Historian exhibits robustness to variations in reference set density and effectively exploits solution redundancy across diverse repair methodologies, provided that semantically equivalent patches exist in the historical record.

\end{tcolorbox}

%% file: tables/rq3-internal.tex
\begin{table*}[t]
\centering
\caption{Leave-One-Tool-Out Cross-Validation}
\label{tab:gemini-comparison}
\footnotesize
\renewcommand{\arraystretch}{1.1}
\setlength{\tabcolsep}{6pt}

\begin{tabular}{l c|cc|cc}
 &  & \multicolumn{2}{c|}{$\toolname^{\textbf{Qwen}}$} & \multicolumn{2}{c}{$\toolname^{\textbf{Gemini}}$} \\
\cline{3-6}

\textbf{Tool} & \textbf{C:O} & \textbf{Acc} & \textbf{Cov(\%)} & \textbf{Acc} & \textbf{Cov(\%)} \\
\hline
ACS         & 29:8   & \cellcolor{blue!15}\textbf{0.61} & 86.5 & 0.59 & 89.2 \\
Arja        & 8:49   & 0.89 & 96.5 & \cellcolor{blue!15}\textbf{0.93} & 98.2 \\
AVATAR      & 17:37  & \cellcolor{blue!15}\textbf{0.91} & 79.6 & 0.90 & 98.1 \\
CapGen      & 9:41   & 0.86 & 92.0 & \cellcolor{blue!15}\textbf{1.00} & 86.0 \\
Cardumen    & 0:9    & \cellcolor{blue!15}\textbf{1.00} & 100 & \cellcolor{blue!15}\textbf{1.00} & 100 \\
DynaMoth    & 1:21   & \cellcolor{blue!15}\textbf{1.00} & 95.5 & 0.96 & 100 \\
FixMiner    & 6:19   & \cellcolor{blue!15}\textbf{0.91} & 88.0 & \cellcolor{blue!15}\textbf{0.91} & 88.0 \\
GenProg     & 1:24   & 0.96 & 96.0 & \cellcolor{blue!15}\textbf{1.00} & 100 \\
HDRepair    & 4:4    & \cellcolor{blue!15}\textbf{1.00} & 87.5 & \cellcolor{blue!15}\textbf{1.00} & 87.5 \\
Jaid        & 32:40  & 0.67 & 94.4 & \cellcolor{blue!15}\textbf{0.76} & 91.7 \\
jGenProg    & 6:33   & 0.79 & 92.3 & \cellcolor{blue!15}\textbf{0.94} & 92.3 \\
jKali       & 4:31   & \cellcolor{blue!15}\textbf{0.85} & 94.3 & \cellcolor{blue!15}\textbf{0.85} & 97.1 \\
jMutRepair  & 2:14   & 0.81 & 100 & \cellcolor{blue!15}\textbf{0.94} & 100 \\
Kali        & 2:36   & 0.95 & 100 & \cellcolor{blue!15}\textbf{0.97} & 100 \\
kPAR        & 2:32   & 0.90 & 91.2 & \cellcolor{blue!15}\textbf{0.94} & 91.2 \\
Nopol       & 6:89   & 0.96 & 95.8 & \cellcolor{blue!15}\textbf{1.00} & 100 \\
RSRepair    & 2:31   & 0.88 & 100 & \cellcolor{blue!15}\textbf{0.97} & 100 \\
SequenceR   & 10:45  & \cellcolor{blue!15}\textbf{0.78} & 94.5 & 0.73 & 92.7 \\
SimFix      & 16:42  & 0.73 & 89.7 & \cellcolor{blue!15}\textbf{0.84} & 96.6 \\
SketchFix   & 5:7    & 0.73 & 91.7 & \cellcolor{blue!15}\textbf{0.75} & 100 \\
SOFix       & 10:1   & \cellcolor{blue!15}\textbf{0.63} & 72.7 & 0.50 & 72.7 \\
TBar        & 7:33   & \cellcolor{blue!15}\textbf{0.94} & 87.5 & 0.87 & 95.0 \\
\hline
\textbf{Avg.}    & - & 0.85 & 92.1 & \cellcolor{blue!15}\textbf{0.88} & 94.4 \\
\textbf{W. Avg.} & - & 84.7 & 92.5 & \cellcolor{blue!15}\textbf{88.4} & 95.0 \\
\textbf{Total}  & 179:646 & - & - & - & - \\
\bottomrule
\end{tabular}

\vspace{1mm}
\parbox{\linewidth}{\centering\footnotesize \textbf{C:O:} Correct:Overfitting,\;
\textbf{Acc:} Accuracy (\%),\;
\textbf{Cov:} Coverage,\;
\textbf{W:} Weighted. \\
\colorbox{blue!15}{Blue} indicates the best accuracy for each tool.}
\end{table*}

%% file: sections/evaluation-p4.tex
\subsection{RQ4: To what extent can Historian serve as a filter to enhance the performance and efficiency of existing APCA tools?}
\label{sec:RQ-complement}


\textbf{Approach:} While the previous research question established Historian’s potential to reduce manual validation workload, we now evaluate its positioning within the broader APCA landscape. Although conventional APCA tools and the Historian framework target distinct operational contexts, probabilistic prediction for novel faults (Scenario 1) versus evidence-based verification for recurring patches (Scenario 2), we investigate whether Historian can serve as a foundation to filter recurring patches for existing techniques. We hypothesize that a two-stage assessment architecture allows both paradigms to operate closer to their intended use-cases: Historian serves as an initial filtering stage that autonomously assigns correctness verdicts to repair attempts with verifiable historical precedent, allowing secondary APCA models to focus their probabilistic inference on the remaining genuinely novel repair attempts.


We evaluate this integrated architecture using three representative state-of-the-art APCA models: ODS \cite{ODS}, Quatrain \cite{quatrain}, and Cache \cite{Cache}. All experiments follow the 22-fold LOTO cross-validation protocol described in RQ3, ensuring that patches generated by each APR tool are assessed only when that tool is excluded from the reference set. For ODS and Quatrain, we re-run the tools on the residual patches that Historian cannot identify a historical evidence. For Cache, we use the data and reported results from the published artifacts. To ensure that our findings are grounded in verified repair outcomes and remain comparable to recent APCA literature, the cross-validation dataset is standardized across all tools, using the patch dataset compiled by Lin et al.~\cite{Cache}. We report results in two parts: (i) the accuracy and F1 uplift achieved by integrating Historian relative to standalone APCA baselines (Table \ref{tab:uplift}), and (ii) a comparison of Historian-integrated pipelines against a broader set of state-of-the-art APCA techniques (Table \ref{tab:comparison}).

\textbf{Results:}

\textbf{Effectiveness of the Integrated Architecture.}
Table \ref{tab:uplift} reports Accuracy and F1 for each standalone APCA model and its corresponding Historian-integrated configuration, along with the absolute improvement achieved by the hybrid pipeline. On average, the hybrid architecture improves absolute accuracy by +21.8\% for ODS, +18.5\% for Quatrain, and +12.6\% for Cache. These improvements indicate that conventional APCA tools are susceptible to misclassifying redundant repair attempts. By identifying redundant repair attempts through historical evidence, Historian isolates these cases, enabling predictive models to focus on a smaller set of potentially novel repairs.

\input{tables/rq6-historian-addtion}

Our analysis reveals that the functional role of Historian within the pipeline is governed primarily by the coverage achieved in the first stage. We identify distinct operational regimes based on the density of available historical evidence. In the Historian-Only Regime, encompassing patches generated by tools such as Cardumen, GenProg, HDRepair, jMutRepair, Kali, Nopol, and RSRepair, coverage under Historian reaches 100\%. In these instances, all repair attempts are assigned to correctness verdicts via a semantic equivalent patch found in the Historical Reference Set, bypassing the secondary APCA tools. The significant accuracy improvements observed here, such as the +62.5 point gain for ODS on Cardumen and +48.4 for ODS on Nopol, reflect the "replacement effect," where standalone probabilistic predictions are substituted by Historian’s high-precision, evidence-based verdicts.


A second Hybrid Regime emerges for patches generated by tools with high but incomplete coverage (between 85\% and 99\%), including ACS, Arja, AVATAR, SimFix, jGenProg, SequenceR, CapGen, FixMiner, and HDRepair. In this regime, APCA tools operate on a small residual subset of patches lacking historical precedent. We observe that weaker baseline APCA tools benefit most from this filtering; for example, Arja repair attempts see substantial gains under ODS (+24.2 points) and Quatrain (+18.0 points), while the stronger Cache model shows less sensitivity to the reduction. Occasional negative deltas, such as the -5.9 point change for SequenceR repair attempts under Cache, indicate that these patches represent atypical edge cases that are inherently more difficult to classify, slightly reducing the measured accuracy on that subset. Finally, we identify a Residual-Sensitive Regime for tools with lower coverage, most notably SOFix (72.7\%). Here, the final accuracy remains highly dependent on the robustness of the secondary classifier. This explains the performance variance for SOFix, which exhibited accuracy declines under ODS but a significant improvement under Cache (+18.2).

\textbf{Contextualization against State-of-the-Art.}
Table~\ref{tab:comparison} contextualizes our integrated hybrid pipelines against prior APCA techniques. The results presented in the left block of the table correspond to the performance values reported in Table II of Zhou et al.~\cite{zhouLeveragingLargeLanguage2024}, which evaluates LLM4PatchCorrect alongside previously proposed APCA approaches on the Defects4J benchmark. We do not independently reproduce these methods. The published values are used solely for contextual comparison. The results in the right block of the table correspond to our hybrid variants (Cache\textsuperscript{Hist.}, ODS\textsuperscript{Hist.}, and Quatrain\textsuperscript{Hist.}) evaluated under the same benchmark and metrics, enabling a directly comparable assessment.

Historian integration substantially enhances APCA performance across APR-generated patch sets, with the strongest improvements observed for GenProg, Nopol, CapGen, Cardumen, and RSRepair, where hybrid configurations often reach 100\% Accuracy. Other patch sets, such as FixMiner, HDRepair, SimFix, Jaid, and kPAR, show moderate gains of 5–13 percentage points over the strongest standalone baseline. In contrast, some APR-generated patch sets remain better addressed by standalone approaches—for example, Arja, jKali, SequenceR, and SOFix—indicating that Historian integration does not uniformly dominate all baselines.

Aggregated statistics further underscore these effects: average Accuracy increases from 63.7–84.4\% (standalone) to 85.5–86.2\% under hybrid variants, with F1 rising from 0.70–0.87 to 0.88. Weighted averages follow a similar pattern, with Accuracy improving from 62.3–84\% to 87.0–87.4\% and weighted F1 reaching 0.90. Overall, Historian integration consistently strengthens APCA performance across most APR-generated patch sets, improving both per-target and aggregate metrics, while the observed exceptions highlight that gains depend on the relative strength of the underlying standalone model.

\input{tables/rq3-comparison}
\begin{tcolorbox}[colback=blue!5,colframe=blue!50!black,title=Summary of RQ4]
Historian complements existing APCA tools by acting as an evidence-based pre-filter in benchmark-centric APR evaluation. When integrated as an initial evidence-based filter, it elevates the accuracy of standalone state-of-the-art models by up to 21.8\%, creating a hybrid assessment pipeline that achieves 86.2\% overall accuracy with 100\% coverage. This integrated architecture assigns correctness verdicts for the majority of repair attempts through traceable historical evidence while reserving probabilistic inference for genuinely novel cases, providing a reliable and efficient assessment solution for APR benchmarking.
\end{tcolorbox}

%% file: tables/rq6-historian-addtion.tex
\begin{table*}[t]
\centering
\caption{Effect of Historian as a Filter}
\label{tab:uplift}

\footnotesize
\renewcommand{\arraystretch}{1.05}
\setlength{\tabcolsep}{2pt}

\begin{tabular}{lr|cc|cc|cc|cc|cc|cc}
\multicolumn{2}{l|}{} 
& \multicolumn{2}{c|}{\textbf{ODS}} 
& \multicolumn{2}{c|}{\textbf{$ODS^{\textbf{\toolname}}$}} 
& \multicolumn{2}{c|}{\textbf{Quatrain}} 
& \multicolumn{2}{c|}{\textbf{$Quatrain^{\textbf{\toolname}}$}} 
& \multicolumn{2}{c|}{\textbf{Cache}} 
& \multicolumn{2}{c}{\textbf{$Cache^{\textbf{\toolname}}$}} \\
\cline{3-14}
\textbf{Target} & \textbf{C:O} 
& \textbf{Acc} & \textbf{F1}
& \textbf{$\Delta$Acc} & \textbf{$\Delta$F1}
& \textbf{Acc} & \textbf{F1}
& \textbf{$\Delta$Acc} & \textbf{$\Delta$F1}
& \textbf{Acc} & \textbf{F1}
& \textbf{$\Delta$Acc} & \textbf{$\Delta$F1} \\
\hline
ACS        & 29:8  & 45.9 & 0.29 & \pos{18.0} & \pos{0.23} & 37.8 & 0.30 & \pos{2.5} & \pos{0.18} & 27.0 & 0.13 & \pos{31.3} & \pos{0.35} \\
Arja       & 8:49  & 68.4 & 0.79 & \pos{24.6} & \pos{0.17} & 63.2 & 0.74 & \pos{29.8} & \pos{0.22} & 82.5 & 0.89 & \pos{1.5} & \pos{0.07} \\
AVATAR     & 17:37 & 64.0 & 0.71 & \pos{19.3} & \pos{0.18} & 72.2 & 0.80 & \pos{13.0} & \pos{0.10} & 74.1 & 0.81 & \pos{14.8} & \pos{0.11} \\
CapGen     & 9:41  & 52.0 & 0.63 & \pos{48.0} & \pos{0.37} & 86.0 & 0.92 & \pos{14.0} & \pos{0.08} & 84.0 & 0.90 & \pos{9.8} & \pos{0.06} \\
\rowcolor{blue!15}
Cardumen   & 0:9   & 37.5 & 0.55 & \pos{62.5} & \pos{0.45} & 77.8 & 0.88 & \pos{22.2} & \pos{0.12} & 66.7 & 0.80 & \pos{33.3} & \pos{0.20} \\
\rowcolor{blue!15}
DynaMoth   & 1:21  & 72.7 & 0.83 & \pos{22.8} & \pos{0.15} & 68.2 & 0.81 & \pos{27.3} & \pos{0.17} & 95.5 & 0.98 & 0.0 & 0.00 \\
FixMiner   & 6:19  & 64.0 & 0.71 & \pos{2.0} & \pos{0.19} & 72.0 & 0.82 & \pos{16.0} & \pos{0.10} & 72.0 & 0.81 & \pos{2.0} & \pos{0.14} \\
GenProg    & 1:24  & 60.0 & 0.75 & \pos{4.0} & \pos{0.25} & 68.0 & 0.80 & \pos{32.0} & \pos{0.20} & 88.0 & 0.94 & \pos{12.0} & \pos{0.06} \\
HDRepair   & 4:4   & 57.1 & 0.67 & \pos{3.4} & \pos{0.22} & 62.5 & 0.73 & \pos{25.0} & \pos{0.16} & 62.5 & 0.73 & \pos{25.0} & \pos{0.16} \\
Jaid       & 32:40 & 48.6 & 0.41 & \pos{22.8} & \pos{0.39} & 68.1 & 0.74 & \pos{8.7} & \pos{0.11} & 61.1 & 0.67 & \pos{13.9} & \pos{0.16} \\
jGenProg   & 6:33  & 69.2 & 0.79 & \pos{2.0} & \pos{0.15} & 82.1 & 0.90 & \pos{7.1} & \pos{0.04} & 84.6 & 0.90 & \pos{7.3} & \pos{0.06} \\
jKali      & 4:31  & 90.3 & 0.95 & \neg{4.6} & \neg{0.03} & 74.3 & 0.85 & \pos{11.4} & \pos{0.07} & 85.7 & 0.92 & 0.0 & 0.00 \\
\rowcolor{blue!15}
jMutRepair & 2:14  & 66.7 & 0.78 & \pos{27.1} & \pos{0.19} & 81.3 & 0.89 & \pos{12.5} & \pos{0.08} & 87.5 & 0.93 & \pos{6.3} & \pos{0.04} \\
\rowcolor{blue!15}
Kali       & 2:36  & 84.2 & 0.91 & \pos{13.2} & \pos{0.08} & 71.1 & 0.83 & \pos{26.3} & \pos{0.16} & 89.5 & 0.94 & \pos{7.9} & \pos{0.05} \\
kPAR       & 2:32  & 50.0 & 0.65 & \pos{38.2} & \pos{0.29} & 64.7 & 0.78 & \pos{26.5} & \pos{0.17} & 79.4 & 0.88 & \pos{8.8} & \pos{0.06} \\
\rowcolor{blue!15}
Nopol      & 6:89  & 51.6 & 0.66 & \pos{48.4} & \pos{0.34} & 71.6 & 0.83 & \pos{28.4} & \pos{0.17} & 89.5 & 0.94 & \pos{1.5} & \pos{0.06} \\
\rowcolor{blue!15}
RSRepair   & 2:31  & 75.8 & 0.85 & \pos{21.2} & \pos{0.13} & 75.8 & 0.86 & \pos{21.2} & \pos{0.12} & 81.8 & 0.89 & \pos{15.2} & \pos{0.09} \\
SequenceR  & 10:45 & 74.5 & 0.83 & \neg{1.8} & \neg{0.02} & 67.3 & 0.79 & \pos{7.7} & \pos{0.04} & 76.4 & 0.85 & \neg{5.9} & \neg{0.06} \\
SimFix     & 16:42 & 60.3 & 0.71 & \pos{2.7} & \pos{0.17} & 69.0 & 0.80 & \pos{12.0} & \pos{0.08} & 69.0 & 0.79 & \pos{12.0} & \pos{0.09} \\
\rowcolor{blue!15}
SketchFix  & 5:7   & 83.3 & 0.86 & \neg{8.3} & \neg{0.09} & 33.3 & 0.33 & \pos{41.7} & \pos{0.44} & 50.0 & 0.63 & \pos{25.0} & \pos{0.14} \\
SOFix      & 10:1  & 72.7 & 0.40 & \neg{36.3} & \neg{0.18} & 45.5 & 0.25 & \neg{9.1} & \neg{0.03} & 27.3 & 0.20 & \pos{18.2} & \pos{0.05} \\
TBar       & 7:33  & 52.5 & 0.61 & \pos{32.5} & \pos{0.30} & 75.0 & 0.85 & \pos{12.5} & \pos{0.07} & 85.0 & 0.91 & \pos{2.5} & \pos{0.01} \\
\hline
\textbf{Avg.} & - & 63.7 & 0.70 & \pos{21.8} & \pos{0.18} & 67.6 & 0.75 & \pos{18.5} & \pos{0.13} & 73.6 & 0.79 & \pos{12.6} & \pos{0.09} \\
\hline
\textbf{W. Avg.} & - & 62.3 & 0.69 & \pos{24.2} & \pos{0.21} & 69.4 & 0.78 & \pos{18.0} & \pos{0.12} & 76.4 & 0.82 & \pos{1.6} & \pos{0.08} \\
\bottomrule
\end{tabular}

\vspace{1mm}
\parbox{\linewidth}{\centering\footnotesize The $\Delta$ columns denote the absolute performance improvement achieved by utilizing Historian as a high-precision first-stage filter. Tools highlighted in \colorbox{blue!20}{blue} indicate that Historian achieves 100\% coverage on their generated patches.}
\end{table*}

%% file: tables/rq3-comparison.tex
\begin{table*}[t]
\centering
\caption{Contextualization Against SOTA APCA}
\label{tab:comparison}
\footnotesize
\setlength{\tabcolsep}{1pt}
\renewcommand{\arraystretch}{1.05}

\begin{tabular}{l r|cc|cc|cc|cc|cc|cc||cc|cc|cc}
\multicolumn{2}{l|}{}  
& \multicolumn{2}{c|}{\textbf{Cache}} 
& \multicolumn{2}{c|}{\textbf{Tian et al.}} 
& \multicolumn{2}{c|}{\textbf{CodeBERT}} 
& \multicolumn{2}{c|}{\textbf{ODS}} 
& \multicolumn{2}{c|}{\textbf{Quatrain}} 
& \multicolumn{2}{c||}{\textbf{LLM4PC}} 
& \multicolumn{2}{c|}{$\mathbf{Cache^{Hist.}}$}
& \multicolumn{2}{c|}{$\mathbf{ODS^{Hist.}}$}
& \multicolumn{2}{c|}{$\mathbf{Quat.^{Hist.}}$} \\
\cline{3-20}
\textbf{Target} & \textbf{C:O} 
& \textbf{Acc} & \textbf{F1}
& \textbf{Acc} & \textbf{F1}
& \textbf{Acc} & \textbf{F1}
& \textbf{Acc} & \textbf{F1}
& \textbf{Acc} & \textbf{F1}
& \textbf{Acc} & \textbf{F1}
& \textbf{Acc} & \textbf{F1}
& \textbf{Acc} & \textbf{F1}
& \textbf{Acc} & \textbf{F1} \\
\hline
ACS        & 29:8  & 27.0 & 0.13 & 40.5 & 0.35 & 43.2 & 0.40 & 45.9 & 0.29 & 37.8 & 0.30 & \cellcolor{blue!15}67.6 & \cellcolor{blue!15}0.57 & 58.3 & 0.48 & 63.9 & 0.52 & 58.3 & 0.48 \\
Arja       & 8:49  & 82.5 & 0.89 & 59.6 & 0.72 & 87.7 & 0.93 & 68.4 & 0.79 & 63.2 & 0.74 & \cellcolor{blue!15}94.7 & \cellcolor{blue!15}0.97 & 93.0 & 0.96 & 93.0 & 0.96 & 93.0 & 0.96 \\
AVATAR     & 17:37 & 74.1 & 0.81 & 74.1 & 0.82 & 75.9 & 0.83 & 64.0 & 0.71 & 72.2 & 0.80 & 85.2 & 0.90 & \cellcolor{blue!15}88.9 & \cellcolor{blue!15}0.92 & 83.3 & 0.89 & 85.2 & 0.90 \\
CapGen     & 9:41  & 84.0 & 0.90 & 84.0 & 0.90 & 82.0 & 0.90 & 52.0 & 0.63 & 86.0 & 0.92 & 80.0 & 0.87 & 93.8 & 0.96 & \cellcolor{blue!15}100.0 & \cellcolor{blue!15}1.00 & \cellcolor{blue!15}100.0 & \cellcolor{blue!15}1.00 \\
Cardumen   & 0:9   & 66.7 & 0.80 & 66.7 & 0.80 & 77.8 & 0.88 & 37.5 & 0.55 & 77.8 & 0.88 & 88.9 & 0.94 & \cellcolor{blue!15}100.0 & \cellcolor{blue!15}1.00 & \cellcolor{blue!15}100.0 & \cellcolor{blue!15}1.00 & \cellcolor{blue!15}100.0 & \cellcolor{blue!15}1.00 \\
DynaMoth   & 1:21  & \cellcolor{blue!15}95.5 & \cellcolor{blue!15}0.98 & 90.9 & 0.95 & \cellcolor{blue!15}95.5 & \cellcolor{blue!15}0.98 & 72.7 & 0.83 & 68.2 & 0.81 & \cellcolor{blue!15}95.5 & \cellcolor{blue!15}0.98 & \cellcolor{blue!15}95.5 & \cellcolor{blue!15}0.98 & \cellcolor{blue!15}95.5 & \cellcolor{blue!15}0.98 & \cellcolor{blue!15}95.5 & \cellcolor{blue!15}0.98 \\
FixMiner   & 6:19  & 72.0 & 0.81 & 68.0 & 0.78 & 80.0 & 0.87 & 64.0 & 0.71 & 72.0 & 0.82 & 84.0 & 0.90 & \cellcolor{blue!15}92.0 & \cellcolor{blue!15}0.95 & 84.0 & 0.90 & 88.0 & 0.92 \\
GenProg    & 1:24  & 88.0 & 0.94 & 68.0 & 0.81 & 96.0 & 0.98 & 60.0 & 0.75 & 68.0 & 0.80 & 92.0 & 0.96 & \cellcolor{blue!15}100.0 & \cellcolor{blue!15}1.00 & \cellcolor{blue!15}100.0 & \cellcolor{blue!15}1.00 & \cellcolor{blue!15}100.0 & \cellcolor{blue!15}1.00 \\
HDRepair   & 4:4   & 62.5 & 0.73 & 62.5 & 0.67 & 75.0 & 0.80 & 57.1 & 0.67 & 62.5 & 0.73 & \cellcolor{blue!15}87.5 & 0.86 & \cellcolor{blue!15}87.5 & \cellcolor{blue!15}0.89 & \cellcolor{blue!15}87.5 & \cellcolor{blue!15}0.89 & \cellcolor{blue!15}87.5 & \cellcolor{blue!15}0.89 \\
Jaid       & 32:40 & 61.1 & 0.67 & 61.1 & 0.70 & 69.4 & 0.76 & 48.6 & 0.41 & 68.1 & 0.74 & 68.1 & 0.72 & 75.0 & 0.83 & 71.4 & 0.80 & \cellcolor{blue!15}76.8 & \cellcolor{blue!15}0.85 \\
jGenProg   & 6:33  & 84.6 & 0.90 & 56.4 & 0.71 & 87.2 & 0.93 & 69.2 & 0.79 & 82.1 & 0.90 & 87.2 & 0.94 & \cellcolor{blue!15}91.9 & \cellcolor{blue!15}0.96 & 89.2 & 0.94 & 89.2 & 0.94 \\
jKali      & 4:31  & 85.7 & 0.92 & 68.6 & 0.81 & \cellcolor{blue!15}94.3 & \cellcolor{blue!15}0.97 & 90.3 & 0.95 & 74.3 & 0.85 & \cellcolor{blue!15}94.3 & \cellcolor{blue!15}0.97 & 85.7 & 0.92 & 85.7 & 0.92 & 85.7 & 0.92 \\
jMutRepair & 2:14  & 87.5 & 0.93 & 68.8 & 0.82 & 87.5 & 0.93 & 66.7 & 0.78 & 81.3 & 0.89 & \cellcolor{blue!15}93.8 & 0.96 & \cellcolor{blue!15}93.8 & \cellcolor{blue!15}0.97 & \cellcolor{blue!15}93.8 & \cellcolor{blue!15}0.97 & \cellcolor{blue!15}93.8 & \cellcolor{blue!15}0.97 \\
Kali       & 2:36  & 89.5 & 0.94 & 76.3 & 0.87 & 92.1 & 0.96 & 84.2 & 0.91 & 71.1 & 0.83 & 92.1 & 0.96 & \cellcolor{blue!15}97.4 & \cellcolor{blue!15}0.99 & \cellcolor{blue!15}97.4 & \cellcolor{blue!15}0.99 & \cellcolor{blue!15}97.4 & \cellcolor{blue!15}0.99 \\
kPAR       & 2:32  & 79.4 & 0.88 & 79.4 & 0.88 & 82.4 & 0.90 & 50.0 & 0.65 & 64.7 & 0.78 & 76.5 & 0.86 & 88.2 & 0.94 & 88.2 & 0.94 & \cellcolor{blue!15}91.2 & \cellcolor{blue!15}0.95 \\
Nopol      & 6:89  & 89.5 & 0.94 & 75.8 & 0.86 & 69.5 & 0.82 & 51.6 & 0.66 & 71.6 & 0.83 & 93.7 & 0.97 & \cellcolor{blue!15}100.0 & \cellcolor{blue!15}1.00 & \cellcolor{blue!15}100.0 & \cellcolor{blue!15}1.00 & \cellcolor{blue!15}100.0 & \cellcolor{blue!15}1.00 \\
RSRepair   & 2:31  & 81.8 & 0.89 & 81.8 & 0.90 & 90.9 & 0.95 & 75.8 & 0.85 & 75.8 & 0.86 & 90.9 & 0.95 & \cellcolor{blue!15}97.0 & \cellcolor{blue!15}0.98 & \cellcolor{blue!15}97.0 & \cellcolor{blue!15}0.98 & \cellcolor{blue!15}97.0 & \cellcolor{blue!15}0.98 \\
SequenceR  & 10:45 & 76.4 & 0.85 & \cellcolor{blue!15}80.0 & \cellcolor{blue!15}0.87 & 61.8 & 0.72 & 74.5 & 0.83 & 67.3 & 0.79 & 72.7 & 0.80 & 70.5 & 0.79 & 72.7 & 0.81 & 75.0 & 0.83 \\
SimFix     & 16:42 & 69.0 & 0.79 & 75.9 & 0.82 & 70.7 & 0.80 & 60.3 & 0.71 & 69.0 & 0.80 & 77.6 & 0.85 & \cellcolor{blue!15}81.0 & \cellcolor{blue!15}0.88 & \cellcolor{blue!15}81.0 & \cellcolor{blue!15}0.88 & \cellcolor{blue!15}81.0 & \cellcolor{blue!15}0.88 \\
SketchFix  & 5:7   & 50.0 & 0.63 & 66.7 & 0.67 & 50.0 & 0.50 & 83.3 & 0.86 & 33.3 & 0.33 & \cellcolor{blue!15}91.7 & \cellcolor{blue!15}0.92 & 75.0 & 0.77 & 75.0 & 0.77 & 75.0 & 0.77 \\
SOFix      & 10:1  & 27.3 & 0.20 & 54.5 & 0.29 & 27.3 & 0.20 & \cellcolor{blue!15}72.7 & \cellcolor{blue!15}0.40 & 45.5 & 0.25 & 54.5 & 0.29 & 45.5 & 0.25 & 36.4 & 0.22 & 36.4 & 0.22 \\
TBar       & 7:33  & 85.0 & 0.91 & 77.5 & 0.86 & \cellcolor{blue!15}87.5 & \cellcolor{blue!15}0.93 & 52.5 & 0.61 & 75.0 & 0.85 & 85.0 & 0.91 & \cellcolor{blue!15}87.5 & 0.92 & 85.0 & 0.91 & \cellcolor{blue!15}87.5 & 0.92 \\
\hline


\textbf{Avg.} & - & 73.6 & 0.79 & 69.9 & 0.77 & 76.5 & 0.82 & 63.7 & 0.70 & 67.6 & 0.75 & 84.4 & 0.87 & \cellcolor{blue!15}86.2 & \cellcolor{blue!15}0.88 & 85.5 & \cellcolor{blue!15}0.88 & 86.1 & \cellcolor{blue!15}0.88 \\
\textbf{Improve.} & - & \pos{12.6} & \pos{.09} & \pos{16.3} & \pos{.11} & \pos{9.7} & \pos{.06} & \pos{22.5} & \pos{.18} & \pos{18.6} & \pos{.13} & \pos{1.8} & \pos{.01} & - & - & \pos{0.7} & - & \pos{0.1} & - \\
\hline
\textbf{W. Avg.} & - & 76.4 & 0.82 & 70.9 & 0.79 & 77.0 & 0.83 & 62.3 & 0.69 & 69.4 & 0.78 & 84.0 & 0.88 & 87.0 & \cellcolor{blue!15}0.90 & 86.5 & \cellcolor{blue!15}0.90 & \cellcolor{blue!15}87.4 & \cellcolor{blue!15}0.90 \\
\textbf{Improve.} & - & \pos{11.0} & \pos{.08} & \pos{16.5} & \pos{.11} & \pos{10.4} & \pos{.07} & \pos{25.1} & \pos{.21} & \pos{18.0} & \pos{.12} & \pos{3.4} & \pos{.02} & \pos{0.4} & - & \pos{0.9} & - & - & - \\

\bottomrule
\end{tabular}

\vspace{1mm}
\parbox{\linewidth}{\centering\footnotesize \colorbox{blue!15}{Blue} indicates the best accuracy or F1 for each tool.}
\end{table*}

%% file: sections/evaluation-p5.tex
\subsection{RQ5. To what extent does solution redundancy persist as the APR benchmarking lifecycle evolves?}
\label{sec:RQ-time}

\textbf{Approach:} We conduct a longitudinal redundancy analysis to investigate the empirical sustainability of evidence-based assessment. Our objective is to determine whether the repetitive nature of the APR benchmarking process, where new tool-generated repair attempts are evaluated against a fixed set of historical bugs (e.g., Defects4J), provides a sustainable mandate for a cumulative knowledge base. We utilize our pre-2020 corpus (450 correct and 8,704 overfitting patches) as the Baseline Historical Reference Set. We then simulate the growth of the APR research ecosystem by chronologically introducing sanitized repair attempts from seven prominent tools released between 2020 and 2024: ARJA-e (2020), Recoder (2021), SelfAPR (2022), Knod (2023), TARE (2023), TransplantFix (2023), and T5APR (2024).

\input{tables/time-systematic}
The availability of peer-reviewed artifacts governed the selection of repair tools for this longitudinal cohort. As summarized in Table~\ref{tab:rq5-tools}, we performed a comprehensive survey of the recent APR landscape. We found that while dozens of tools have been proposed since 2020, data-release practices remain inconsistent across the community. Many recent methodologies, such as AlphaRepair, ChatRepair, and RAPGen, provide only Correct patches, lacking the labeled Overfitting patches required for a comprehensive assessment study. Other studies, such as DEAR and GAMMA, either do not separate overfitting patches from plausible ones or do not provide patch contents. To ensure that our findings are grounded in original, expert-validated outcomes rather than in our own re-execution, we included only tools that provide a complete and reproducible set of repair artifacts.
\input{tables/time-systematic-dissection}
To ensure the statistical rigor of this analysis, we implemented a three-stage Patch Deduplication Pipeline, detailed in Table~\ref{tab:dedup_pipeline}. We first collected and corrected the raw artifacts to ensure that they could be successfully applied to the project using \texttt{git apply}, resulting in an initial set of 1,431 Correct and 4,082 Overfitting patches. Following the scope established in Section ~\ref{sec:patch-dataset}, we filtered this set to include exclusively single-method repair attempts, facilitating a consistent comparison across localized and holistic code representations. Finally, we performed a rigorous de-duplication to ensure that each repair attempt for a specific bug was unique within its tool-generated set. This sanitization process yielded a finalized evaluation corpus of 862 Correct and 2,232 Overfitting patches.


We identify redundancy within this corpus using a deterministic, three-stage cascading detection pipeline (Exact, Token, and AST-based matching). Exact Match check for textual identity in modified method bodies; Token-based Match using SourcererCC (similarity threshold of 1.0) to identify Type-1 and Type-2 clones; and AST-based Detection stage employing tree-differencing to capture structural identity despite identifier renaming. This cascading approach establishes an objective, conservative lower bound for solution redundancy that is independent of model-specific reasoning bias. To simulate the real-world accumulation of knowledge, we evaluate the tools sequentially: after a tool’s repair attempts are analyzed, they are integrated into the reference set before evaluating the subsequent tool in the timeline. We categorize each identified redundant patch by its match origin: Baseline Only (matching the pre-2020 record), Added Tools Only (matching exclusively within the post-2020 cohort), or Both.

\textbf{Results:}

\textbf{Convergence in the Repair Solution Space.}
The results of our longitudinal analysis, detailed in Table~\ref{tab:patch_analysis}, characterize solution recurrence as a persistent and significant property of the APR benchmarking lifecycle. Analysis of the 862 validated correct patches identifies a significant degree of redundancy, with 341 repair attempts (39.6\%) independently rediscovering established fixes. For the earliest tool in the cohort, ARJA-e (2020), redundancy was driven exclusively by the original pre-2020 record, with 50.0\% (18/36) of its patches matching the baseline. However, as the timeline progressed, we observed a marked shift toward a more diverse evidence base. For the most recent tool, T5APR (2024), 124 out of 415 correct repair attempts were redundant. Notably, 98.4\% of these matches originated from the post-2020 record (54 "Added Only" and 68 "Both"). This confirms that as the ecosystem matures, new repair methodologies increasingly build upon or rediscover the repair logic established by their immediate predecessors.
\input{tables/redundancy}

The prevalence of the "Both" category, comprising 166 patches (19.3\% of the total correct corpus), is particularly revealing of the field's trajectory. These patches represent repair logic originally found in the baseline that was independently rediscovered by multiple modern tools. In the case of Knod (2023), nearly half of its correct patches (31/69) fall into this category. These results indicate that the APR field is not diverging toward infinite novelty; rather, it is mapping a finite and recurring solution space, which directly enhances the long-term utility of a Historical Reference Set.

\textbf{Shared Failures in Overfitting.}
While the redundancy rate for overfitting repair attempts is lower (14.9\% overall), the results provide compelling existence proof for recurring logic errors. Out of 2,232 overfitting attempts, we identified 333 redundant instances, with recurrence being most pronounced in high-volume recent tools such as T5APR (2024). In the T5APR dataset, 236 overfitting repair attempts matched the historical record, with the majority (n=175) found exclusively in the "Added Tools" cohort. This suggests that recent repair methodologies frequently yield to identical flawed repair logic. This recurrence justifies Historian’s design choice to incorporate validated logic errors into its reference set, allowing the framework to filter out redundant flaws with factual certainty.

\textbf{Quantifying the Workload Reduction Potential.}
This longitudinal analysis provides a concrete quantification of the manual labor that can be bypassed through an evidence-based approach. Across the 3,094 tool-generated patches studied in this cohort, we identified 674 instances (21.8\%) of structural redundancy relative to the evolving historical record. In the context of APR benchmarking, these results represent a verified conservative lower bound for workload reduction. Because our detection pipeline utilizes deterministic structural tools, it is blind to the deeper semantic equivalences that Historian’s reasoning-based logic is designed to capture. Consequently, the 21.8\% figure represents the absolute minimum volume of manual validation that is methodologically redundant; the true potential for automation is likely significantly higher.

Crucially, for state-of-the-art tools, the data indicates that nearly one-fifth of the manual validation workload is spent re-evaluating repair logic already established in the cumulative record. These findings prove that Historian is not a static system, but a sustainable scientific infrastructure. By proving that the APR solution space is convergent rather than divergent, we demonstrate that the "automation opportunity" expands alongside the ecosystem. This provides the opportunity to systematically preserve and reuse prior manual validation efforts.




\begin{tcolorbox}[colback=blue!5,colframe=blue!50!black,title=Summary of RQ5]
Solution redundancy is a persistent property of the APR benchmarking lifecycle, with 39.6\% of recent correct patches independently rediscovering established fixes. We establish a 21.8\% conservative lower bound for aggregate workload reduction based on structural redundancy alone. These results prove that the effectiveness of the evidence-based assessment is \textbf{cumulative}: as the community archives new repair attempts, the likelihood of encountering semantically redundant patches increases, providing a sustainable, high-precision solution for mitigating the manual validation bottleneck as the repair ecosystem matures.
\end{tcolorbox}

%% file: tables/time-systematic.tex
\begin{table}[t]
\centering
\caption{List of APR tools used in this study}
\label{tab:rq5-tools}
\footnotesize 
\setlength{\tabcolsep}{4pt}
\begin{tabular}{l l}
\textbf{Status} & \textbf{Study} \\
\midrule
No Overfitting Patches & Circle~\cite{yuan2022circle} (2022), AlphaRepair~\cite{alpharepair} (2022), ChatRepair~\cite{chatrepair} (2024), \\
& Coconut~\cite{coconut} (2020), Xia et al.~\cite{codex} (2020), Cure~\cite{cure} (2021), \\
& DLFix~\cite{li2020dlfix} (2020), FitRepair~\cite{fitrepair} (2023), ITER~\cite{ye2024iter} (2024), \\
& Liana~\cite{liana} (2023), MCRepair~\cite{kim2023mcrepair} (2023), MulPOR~\cite{mul} (2024), \\
& RAPGen~\cite{wang2023rapgen} (2023), Repilot~\cite{repilot} (2023), RewardRepair~\cite{rewardrepair} (2022), \\
& Tenure~\cite{tenure} (2023), ThinkRepair~\cite{yin2024thinkrepair} (2024) \\
\midrule
No Clean Patches & DEAR~\cite{li2022dear}, TRANSFER~\cite{transfer}, GAMMA~\cite{zhang2023gamma} (2023) \\
\midrule
Included & ARJA-e~\cite{arjae} (2020), Recoder~\cite{recoder} (2021), SelfAPR~\cite{ye2022selfapr} (2022), \\
& Knod~\cite{jiang2023knod} (2023), TARE~\cite{zhu2023tare} (2023), TransplantFix~\cite{yang2022transplantfix} (2023), \\
& T5APR~\cite{gharibi2024t5apr} (2024)\\
\end{tabular}
\end{table}




%% file: tables/time-systematic-dissection.tex
\begin{table}[t]
\centering
\caption{Patch Deduplication Pipeline Results}
\label{tab:dedup_pipeline}
\footnotesize
\setlength{\tabcolsep}{5pt}
\renewcommand{\arraystretch}{1.05}

\begin{tabular}{l|cc|cc|cc}
\multirow{2}{*}{\textbf{Tool}} 
& \multicolumn{2}{c|}{\textbf{Cleaned}} 
& \multicolumn{2}{c|}{\textbf{Single Methods}} 
& \multicolumn{2}{c}{\textbf{Deduplicated}} \\
\cline{2-7}
& \textbf{Correct} & \textbf{Overfit} 
& \textbf{Correct} & \textbf{Overfit}
& \textbf{Correct} & \textbf{Overfit} \\
\hline
ARJA-e & 39 & 67 & 36 & 62 & \cellcolor{blue!15}36 & \cellcolor{blue!15}62 \\
Recoder & 85 & 5 & 77 & 5 & \cellcolor{blue!15}77 & \cellcolor{blue!15}5 \\
SelfAPR & 97 & 516 & 91 & 498 & \cellcolor{blue!15}69 & \cellcolor{blue!15}428 \\
Knod & 71 & 14 & 69 & 14 & \cellcolor{blue!15}69 & \cellcolor{blue!15}14 \\
TARE & 152 & 141 & 151 & 141 & \cellcolor{blue!15}124 & \cellcolor{blue!15}134 \\
TransplantFix & 113 & 161 & 111 & 160 & \cellcolor{blue!15}72 & \cellcolor{blue!15}123 \\
T5APR & 874 & 3178 & 811 & 2792 & \cellcolor{blue!15}415 & \cellcolor{blue!15}1466 \\
\hline
\textbf{Total} & \textbf{1431} & \textbf{4082} & \textbf{1346} & \textbf{3672} & \cellcolor{blue!15}\textbf{862} & \cellcolor{blue!15}\textbf{2232} \\
\bottomrule
\end{tabular}

\vspace{1mm}
\parbox{\linewidth}{\centering\footnotesize C = Correct patches, O = Overfitting patches. \colorbox{blue!15}{Blue} highlights the final number of patches.}

\end{table}

%% file: tables/redundancy.tex
\begin{table*}[t]
\centering
\caption{Patch Classification Analysis Across APR Tools}
\label{tab:patch_analysis}
\footnotesize
\setlength{\tabcolsep}{4pt}
\renewcommand{\arraystretch}{1.05}
\begin{tabular}{l|rr|rrr|rrr}
\multirow{2}{*}{\textbf{Tool}} 
& \multicolumn{2}{c|}{\textbf{Total Patches}} 
& \multicolumn{3}{c|}{\textbf{Correct Patches}} 
& \multicolumn{3}{c}{\textbf{Overfitting Patches}} \\
\cline{2-9}
& \textbf{Correct} & \textbf{Overfit} 
& \textbf{Baseline} & \textbf{Added} & \textbf{Both}
& \textbf{Baseline} & \textbf{Added} & \textbf{Both} \\
\hline
ARJA-e & 36 & 62 & 18 & 0 & 0 & 22 & 0 & 0 \\
Recoder & 77 & 5 & 28 & 0 & 7 & 3 & 0 & 0 \\
SelfAPR & 69 & 428 & 9 & 4 & 22 & 6 & 6 & 8 \\
Knod & 69 & 14 & 6 & 8 & 31 & 0 & 1 & 1 \\
TARE & 124 & 134 & 9 & 23 & 29 & 17 & 16 & 3 \\
TransplantFix & 72 & 123 & 9 & 5 & 9 & 12 & 1 & 1 \\
T5APR & 415 & 1466 & 2 & 54 & 68 & 27 & 175 & 34 \\
\hline
\textbf{Total} & \textbf{862} & \textbf{2232} & \textbf{81} & \textbf{94} & \textbf{166} & \textbf{87} & \textbf{199} & \textbf{47} \\
\end{tabular}
\vspace{1mm}
\end{table*}

%% file: sections/discussion.tex
\section{Discussion}\label{sec:discussion}

The results of our comprehensive evaluation provide a rigorous validation of the evidence-based assessment paradigm. By achieving high automation coverage and precision across diverse repair tool generations, Historian demonstrates that the inherent redundancy within the APR benchmarking lifecycle provides a sustainable opportunity for workload reduction. In this section, we reflect on the fundamental shift from prediction to verification, characterize the operational limits of the framework, and discuss the implications of our findings for the broader research community.

\textbf{From Probabilistic Prediction to Fact-Based Verification.} A central contribution of this work is the re-characterization of assessment within the benchmark-centric research lifecycle. While conventional classification models utilize probabilistic inference to estimate correctness, our longitudinal analysis proves that the benchmarking lifecycle is governed by convergence, where repair attempts repeatedly rediscover patches. In this setting, we argue that assessment should be treated as a task of historical verification rather than estimation. Our results establish that when semantically equivalent precedents are identified, the resulting verdicts achieve near-perfect precision. This transforms the "black-box" nature of automated assessment into a transparent process where the majority of verdicts are directly traceable to specific historical evidence. By anchoring the assessment in a Historical Reference Set, we facilitate a more reliable and auditable evaluation of repair methodologies.

\textbf{Influence of Evidence Density on Assessment Reliability.}
Our findings reveal a critical technical insight into the efficiency of evidence-based assessment: the primary driver of performance is the specificity of the semantic link rather than the absolute density of the historical record. We observed that Historian maintained high precision even for bugs with sparse repair history, confirming that the identification of a single unambiguous semantically equivalent precedent is often sufficient to render a definitive verdict on a repair attempt. This robustness ensures that the framework remains effective during the early stages of benchmark growth, where evidence is naturally constrained.


\textbf{Division of Labor.}
The success of the hybrid pipelines in RQ4 establishes Historian as a foundational component for patch assessment. By acting as an evidence-based filter, Historian automates the assessment of redundant repair attempts, allowing classification models to focus their probabilistic inference exclusively on the genuinely novel residual cases. This architectural division of labor ensures that each paradigm operates within its optimal context: Historian provides evidence-based verification for recurring patches, while classification models provide predictive inference for unique patches. The resulting performance convergence observed across diverse classifiers suggests that evidence-based filtering is a critical driver of quality. This suggests a strategic shift for the community: rather than striving for marginal gains in standalone probabilistic accuracy, the community should prioritize the accumulation and documentation of diverse, validated repair outcomes. As the APR research community continues to archive more diverse repair signatures, the empirical scope and reliability of this integrated ecosystem will naturally expand.


\textbf{Long-term Sustainability.}
Our longitudinal study proves that this approach is inherently sustainable. The persistent and growing redundancy observed across tool generations (reaching 61.5\% for recent tools) confirms that the "repair solution space" for common benchmark bugs is stable. This ensures that the Historical Reference Set will remain a valid anchor for future research, transforming the assessment bottleneck into a cumulative knowledge-sharing process that becomes increasingly powerful as the ecosystem matures.

%% file: sections/threats.tex
\section{Threats to Validity}
\label{sec:threats}

\textbf{Internal Validity.} Our work relies on two sources of labels. First, the historical correctness labels for our reference set are drawn from prior studies and may contain errors. We mitigate this by aggregating data from five established, peer-reviewed collections and believe our majority voting mechanism offers some resilience to isolated label noise. Second, our manual clone-type annotations (for our motivating study and RQ1) could be biased. We addressed this by implementing a rigorous, blinded labeling protocol conducted independently by two authors. The high inter-annotator agreement (Cohen’s $\kappa = 0.96$) confirms the reliability of this oracle baseline, with all initial disagreements resolved through consensus discussion.
   
\noindent\textbf{Construct Validity.} Our framework utilizes LLM-based judgments of similarity as an operational proxy for semantic equivalence. This approach is susceptible to the reasoning noise and fallibility inherent in current LLMs. We implement three primary mitigations to ensure the reliability of this construct: (i) we utilize a deterministic temperature setting (0.1) to minimize stochastic variance in model outputs; (ii) we utilize a robust two-stage response parsing pipeline to accurately extract relationship labels from unstructured reasoning, preserving model integrity while ensuring structured input for the subsequent logic; and (iii) we perform a systematic sensitivity analysis across diverse models and prompts (RQ2) to identify the configurations that maximize assessment precision. Furthermore, our oracle experiment in RQ1 establishes the analytical upper bound of our logic in isolation from LLM-induced noise, validating the soundness of the underlying inference rules.

\noindent\textbf{External Validity.} The generalizability of our findings is subject to three primary constraints. First, our experiments are conducted exclusively on the Java ecosystem. While the design of Historian is fundamentally language-agnostic, we selected Java as the evaluation target because it represents the most mature and extensively benchmarked ecosystem in the APR community. Second, our intensive manual analyses in RQ1 utilized repair attempts generated by the TBar tool. While TBar is a state-of-the-art repair system whose output is indicative of canonical repair strategies, the generalizability of these oracle results to other tool architectures remains to be confirmed. We mitigate this by conducting a large-scale cross-tool evaluation (RQ3) and a longitudinal redundancy analysis (RQ5) across 22 diverse repair methodologies. The high performance observed in these broader experiments suggests that the evidence-based paradigm is robust across heterogeneous repair signatures. Finally, our results with a specific suite of LLMs may not generalize to future models; however, the consistent performance scaling observed when moving from open-source architectures to Gemini 2.0 Flash indicates that the framework is well-positioned to benefit from continued advancements in model reasoning capacity.

%% file: sections/related.tex
\section{Related Work}
\label{sec:related_work}
\subsection{Automated Patch Correctness Assessment}
The primary paradigm in Automated Patch Correctness Assessment (APCA) involves abstract predictive models~\cite{yang2023pccsurvey, le2019reliability, zhangAPPTBoostingAutomated2024}. These tools aim to predict a patch's correctness by learning from various features. Some analyze static source code, using handcrafted AST features, such as ODS~\cite{2-ye2021automated}, or code embeddings from models like CodeBERT~\cite{feng2020codebert,4-tian2020evaluating}, or context-aware code change embeddings~\cite{Cache}, while others analyze patch features~\cite{bennettAutomaticallyGeneratedPatches2022}. Quatrain~\cite{quatrain} correlates bug reports with patch descriptions to detect patch correctness. Others incorporate dynamic features from program execution, such as test generation in Patch-SIM~\cite{patchsim}, RGT~\cite{1-ye2021automated}, and xTestCluster~\cite{martinezTestbasedPatchClustering2024} or invariant detection in Invalidator~\cite{7-le2023invalidator} or execution traces in the recent LLM4PatchCorrect~\cite{zhouLeveragingLargeLanguage2024}. Historian's primary distinction is its paradigm shift away from abstract prediction. Instead of training an opaque model to learn "correctness", our framework makes traceable, evidence-based comparisons against a concrete set of historical examples, a novel approach designed to directly leverage the patch redundancy we identified in our motivating study.


The concept of using similarity has been explored before. For instance, tools like Shibboleth~\cite{10-ghanbari2022patch} and MIPI~\cite{12-phung2022identifying} use syntactic or lexical similarity as one signal among many within a predictive model. Historian fundamentally extends this idea in two ways. First, our multi-reference strategy compares against an entire ecosystem of both Correct and Overfitting validated patches, allowing us to handle the high solution diversity we observed. Second, we use powerful LLMs for deep semantic comparison, moving beyond superficial similarity to understand a patch's functional intent. This combination of a multi-reference knowledge base and deep semantic reasoning is the core novelty of our evidence-based framework.

\subsection{Code Clone Detection}
Code clone detection identifies functionally similar code fragments, traditionally categorized into four types~\cite{roy2007survey}. Early methods used text, token, tree, or graph-based analyses~\cite{ducasse1999language,jiang2007deckard, kamiya2002ccfinder}. Recent work explores LLMs for clone detection: Martinez-Gil et al.\cite{martinez-gilSourceCodeClone2024} examine unsupervised methods, Moumoula et al.\cite{moumoulaCrosslingualCodeClone2024} frame cross-lingual detection as classification tasks, Ma et al.\cite{maUnveilingCodePreTrained2024} compare pre-trained models against coding LLMs, Nashaat et al.\cite{nashaatEnhancedTransformerBasedFramework2025} introduce CloneXformer framework, Khajezade et al.\cite{khajezadeInvestigatingEfficacyLarge2024} investigate zero-shot ChatGPT prompts, Zhang et al.\cite{zhang2024assessingCodeClone} use few-shot prompts on GPT models, and Dou et al.~\cite{dou2023towards} systematically evaluate LLMs for clone detection. While these works focus on developing or evaluating clone detection itself, our paper is the first to apply modern, LLM-based semantic clone detection as the core reasoning engine for our new evidence-based patch assessment paradigm. We demonstrate how this powerful technique, when combined with our multi-reference historical set and principled inference logic, can effectively and scalably determine patch correctness.

\subsection{Prompt Engineering}
Recent research highlights the critical role of prompt engineering in harnessing LLMs for automated software engineering tasks~\cite{fan2023large,ouedraogo2024llms,vul,llmshot,solid,ouedraogo2024test,wang2024software,koyuncu2025exploring}. Basic techniques like zero-shot and few-shot learning~\cite{brown2020language,kojima2022large} have been widely used. More advanced strategies, including Chain-of-Thought reasoning~\cite{wei2022chain}, self-consistency~\cite{wang2022self}, and Tree-of-Thoughts~\cite{yao2023tree}, improve model interpretability and performance by eliciting multi-step reasoning. Our work builds upon this body of knowledge. We do not propose a new prompting technique, but instead conduct a rigorous, empirical evaluation of diverse prompt strategies (e.g., Simple, Reasoning-based, Integrated) specifically for the task of determining semantic equivalence between code patches. The findings from this systematic study (RQ2) provide concrete, actionable guidance on the optimal ways to prompt LLMs for our new evidence-based patch assessment paradigm.

%% file: sections/conclusion.tex
\section{Conclusion}
\label{sec:conclusion}

In this work, we addressed the critical bottleneck of patch correctness assessment in Automated Program Repair, a process currently hampered by the methodological inefficiency of repeatedly evaluating semantically redundant solutions. Our foundational empirical study quantified a crucial duality in the patch landscape: a high degree of redundancy (\textasciitilde39\% of unique repair attempts are syntactic clones) alongside significant solution diversity (\textasciitilde65\% of bugs possess multiple correct fixes). This duality motivated a fundamental paradigm shift from opaque, probabilistic prediction toward traceable, evidence-based verification. We presented Historian, the first framework designed to treat the ecosystem of validated repairs as a dynamic and verifiable Historical Reference Set.

Our comprehensive evaluation confirmed the transformative potential of this paradigm through five research questions. An oracle study first demonstrated that our two-stage Evidence-Based Inference Logic possesses a 100\% precision ceiling, ensuring that verdicts grounded in historical evidence are infallible. In a subsequent large-scale, 22-fold leave-one-tool-out evaluation, the framework demonstrated its practical utility by autonomously rendering verdicts for 95.0\% of the aggregate evaluation corpus with 88.4\% accuracy. We further established that Historian can act as an evidence-based pre-filter for the broader ecosystem, elevating the accuracy of standalone state-of-the-art APCA models by up to 21.8\% in hybrid pipelines. Finally, our longitudinal analysis of repair attempts produced across a five-year period (2020–2024) proved that redundancy is a persistent property of the benchmarking lifecycle, confirming that the framework’s utility is inherently cumulative: as the community archives more validated repairs, the automation potential of the evidence-based paradigm expands.
 
By automating the assessment of the large volume of redundant repair attempts, Historian fundamentally changes the economics of APR research. It allows human experts and classification-based tools to bypass repetitive workloads and focus their finite cognitive and computational resources exclusively on the genuinely novel repairs that advance the state of the art. This work establishes evidence-based assessment as a pragmatic, sustainable, and trustworthy new paradigm, ensuring that the cumulative knowledge of the APR community is systematically leveraged to make the program repair assessment more efficient, transparent, and reproducible.

\clearpage
